\documentclass[a4paper, reprint, showkeys, nofootinbib, twoside]{revtex4-1} 

\usepackage{amsmath, amssymb, amsfonts}
\usepackage[english]{babel}
\usepackage[utf8]{inputenc}
\usepackage[colorinlistoftodos, color=green!40, prependcaption]{todonotes}
\usepackage[pdftex, pdftitle={Article}, pdfauthor={Author}]{hyperref} 
\usepackage{amsthm}
\usepackage{mathtools}
\usepackage{physics}
\usepackage{xcolor}
\usepackage{graphicx}
\usepackage[left=23mm,right=13mm,top=35mm,columnsep=15pt]{geometry} 
\usepackage{adjustbox}
\usepackage{placeins}
\usepackage[T1]{fontenc}
\usepackage{lipsum}
\usepackage{csquotes}

\usepackage{amsmath}
\usepackage{subcaption}
\usepackage{tabularx}
\usepackage{enumerate}
\usepackage{ulem} 
\usepackage{mathtools}
\usepackage{eqparbox}
\usepackage{parskip}

\newcommand{\g}[2]{G_{\boldsymbol{#1}}^{(#2)}}
\newcommand{\G}[2]{G_{\boldsymbol{#1}(#2)}}

\newcommand{\dis}[2]{#1^{(#2)}}
\newcommand{\bs}[1]{\boldsymbol{#1}}

\usepackage{babel}

\begin{document}
\preprint{APS/123-QED}

\title{Modeling diffusion in networks with communities: a multitype branching process approach}

\author{Alina Dubovskaya}
\email[Correspondence email address: ]{Alina.Dubovskaya@ul.ie}
\author{Caroline B. Pena}
\author{David J.P. O'Sullivan}
\affiliation{University of Limerick, Limerick, Ireland
}

\date{\today}

\begin{abstract}  
    The dynamics of diffusion in complex networks are widely studied to understand how entities, such as information, diseases, or behaviors, spread in an interconnected environment. Complex networks often present community structure, and tools to analyze diffusion processes on networks with communities are needed. In this paper, we develop theoretical tools using multi-type branching processes to model and analyze diffusion processes, following a simple contagion mechanism, across a broad class of networks with community structure. We show how, by using limited information about the network --- the degree distribution within and between communities --- we can calculate standard statistical characteristics of propagation dynamics, such as the extinction probability, hazard function, and cascade size distribution. These properties can be estimated not only for the entire network but also for each community separately. 
    Furthermore, we estimate the probability of spread crossing from one community to another where it is not currently spreading. We demonstrate the accuracy of our framework by applying it to two specific examples: the Stochastic Block Model and a log-normal network with community structure. We show how the initial seeding location affects the observed cascade size distribution on a heavy-tailed network and that our framework accurately captures this effect.
    \end{abstract}

\maketitle


\section{Introduction and Background}
Understanding the dynamics of spreading phenomena across complex networks is crucial in today's interconnected world~\cite{siddiqui2016social}. Various forms of spread, whether they involve diseases, innovations, behaviors, or information, have a profound impact on human societies. The way diseases spread can determine public health outcomes~\cite{meagher_assessing_2022,fyles_using_2021}, the way innovations diffuse can shape economic development~\cite{mccullen2013multiparameter,peres2014impact}, and the way behaviors and ideas propagate can influence social and political landscapes~\cite{siddiqui2016social, persily_social_2020,juul2019hipsters,aral_distinguishing_2009,clark_role_2022, johnson_online_2020, noauthor_infodemics_nodate, beguerisse-diaz_who_2017}. It is, therefore, crucial to understand propagation dynamics among agents connected via complex networks and what factors accelerate or slow down the spread. Consequently, many efforts have been devoted to developing mathematical tools to analyze propagation on networks \cite{boguna_absence_2003, miller_percolation_2009, porter_dynamical_2016}. In the study of diffusion processes, we often deal with the process unfolding on a complex network of connections, where the structure of the network becomes crucial in determining the dynamics. These networks can be complex, but there are ways to measure and summarize them, with many characteristics being well-known. 

Community structure is one of the most discussed structural properties of social and physical contact networks. Previous studies have highlighted the importance of community structure roles in explaining the spread of social contagions via reinforcement and homophily using both empirical and statistical analyses 
\cite{weng_virality_2013,zhou_survey_2022}. Some have used communities as a feature to predict cascade size using machine learning and statistical methods \cite{hoang_gpop_2017, weng_predicting_2014}. Community structure in real-world networks has been shown to have a substantial impact on percolation and epidemic spreading processes, where the community structure can both benefit and slow down the diffusion processes \cite{stegehuis_epidemic_2016}. Additionally, mathematical models for spreading processes have also been developed, where probability-generating functions (PGFs) are used to capture the network and community structure, and then standard ordinary differential equations compartmental models are applied to model the diffusion processes \cite{li_sir_2018}. However, the literature is scarce when it comes to theoretical studies of propagation dynamics on a network with community structure using branching processes~\cite{li_influence_2018,gomez-rodriguez_inferring_2012,peixoto_network_2019}. One notable exception is Brummitt et al.'s~\cite{brummitt_suppressing_2012} work on modeling interconnected electricity grids, which showed the impact of sparsely connected electricity networks on the occurrence of blackouts of various sizes. 

The branching processes framework allows us to estimate the distributional properties of the process. Instead of focusing on the average behavior of cascades, which has been the main focus of many previous works, in this paper, we are interested in estimating distributions of the typical statistical characteristics of the cascade dynamics, e.g., cascade size distribution and extinction time distribution. Branching process approximation has previously proven effective in studying diffusion across local tree-like networks~\cite{gleeson_branching_2021}, as well as on both homogeneous clique type networks and networks with heterogeneous degree and clustering~\cite{keating_multitype_2022, Keating_Gleeson_OSullivan_2023}. These works laid down a theoretical formalism that we can extend to networks with community structure. The introduction of a community structure adds additional complexity to the analysis of cascades. Spreading may halt within a community but can later be reintroduced from another community, which, for example, affects the calculation of the probability that diffusion stops (extinction probability) within a community.  This complexity requires the development of new mathematical methods to describe diffusion processes across communities. 

In this paper, we develop theoretical tools to model and analyze propagation dynamics on a broad class of networks with community structure for a simple contagion mechanism that follows the Independent Cascade Model (ICM)~\cite{kempe_maximizing_2003,lerman_information_2016}. In the ICM model, each exposure of a susceptible individual to a disease (or piece of information) by an infected friend (or followed person) results in an independent chance of disease transmission (information adoption). We use a multi-type branching process approximation~\cite{caswell_matrix_2002} and probability-generating functions to describe the spread under the ICM. We derive analytical expressions for fundamental characteristics of spreading phenomena, such as the probability of extinction, hazard function, and cascade sizes distribution. Furthermore, we derive specific features associated with cross-community diffusion, such as the introduction probability --- the probability of information (or disease) (re)introduction to a community where it is not currently spreading. We demonstrate the effectiveness of our analysis initially on the Stochastic Block Model and then extend it to a general locally tree-like network with communities with known degree distributions.

The rest of this paper is structured as follows. We begin by describing the model in Sec.~\ref{sec:model_description}, where we outline the assumptions we make about the underlying network and what is required for our method to be applicable. In Sec.~\ref{sec:pgf_framework}, we explain our approach using the Stochastic Block Model, a simple network with community structure. We demonstrate how we can simultaneously track cascade dynamics in multiple communities using multivariate probability generation functions. We derive extinction probabilities, hazards, and cascade size distributions separately for each community as well as for the entire network. In Sec.~\ref{sec:Custom_network}, we extend the analysis from the previous section to a broad class of local-tree networks with communities. We then consider a typical example of a heavy-tailed network with communities and discuss how the community structure affects the cascade size distribution. Finally, Sec.~\ref{sec:Conclusion} provides our concluding remarks and outlines areas for future research.

\section{Conceptual model description of the spreading process on networks with communities}\label{sec:model_description}

\begin{figure*}[t]
  \centering
  \includegraphics[width=0.9\textwidth]{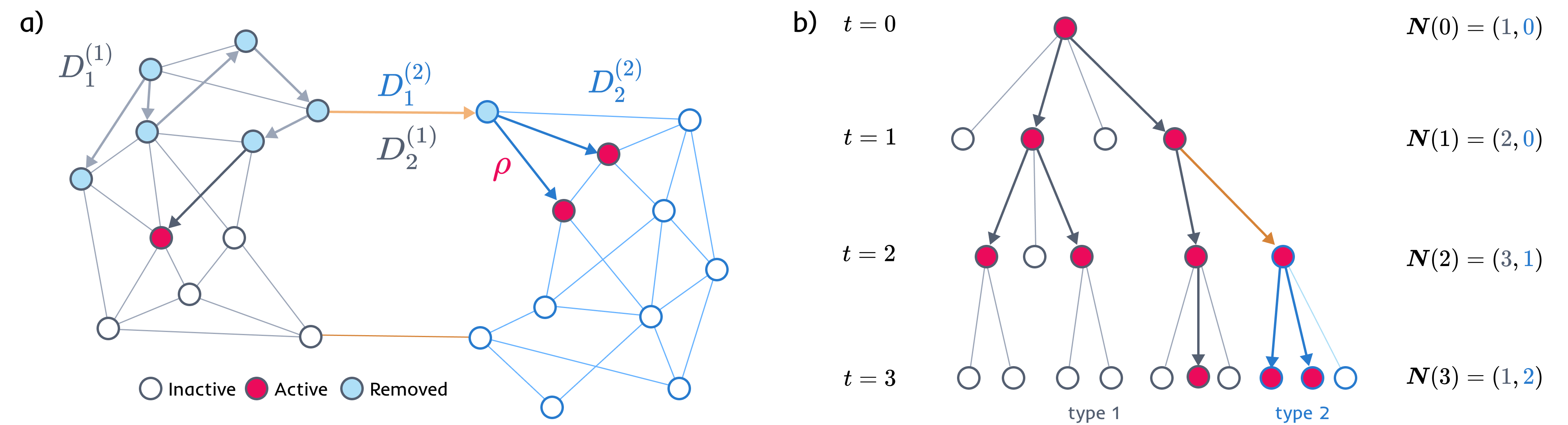}
  \caption{a) Schematic illustration of the Independent Cascade Model (ICM) on a community-based network. The network consists of two communities
  each described by its internal degree distributions and distribution of edges between communities. Here $D_1^{(1)}$ and $D_2^{(2)}$ are random variables for the nodes degree inside community 1 and 2 respectively; $D_2^{(1)}$ and $D_1^{(2)}$ are random variables for a nodes' degree between-community (they are the same in the case of undirected networks considered here). At each time step, active nodes attempt to activate neighbors with probability $\rho$, then become inactive (or ``removed''). The spread starts in the first community initially and later propagates to the second community. b) Schematic of the multi-type branching process approximating ICM on a community-based network. We track two types of offspring: ``type 1'' for active nodes in community 1, and ``type 2'' for active nodes in community 2. Here \mbox{$\boldsymbol{N}(t)=(N_1(t),N_2(t))$} tracks the number of active nodes of type 1 ($N_1(t)$) and type 2 ($N_2(t)$) at each time step.}
  \label{fig:model}
\end{figure*}

In this paper, we study the diffusion process following the Independent Cascade Model on community-based networks.  
The schematic of the process along with the type of network we study is illustrated in Fig.~\ref{fig:model}a). The \textit{Independent Cascade Model (ICM)}  defines nodes of a network as being in one of three states: inactive, active, and removed. Active nodes independently attempt to activate their inactive neighbors with probability $\rho$, transitioning to the removed state afterward. Once the node is in the removed state, it cannot be reactivated. The process begins with a randomly selected active node from community 1 and terminates when no new active nodes are generated. 
We consider diffusion that can be modeled as a \textit{subcritical branching process}, which is certain to terminate and only spread to a fraction of the possible nodes in a network. Such processes are interesting as many of the social or disease spreading processes we wish to capture are subcritical, where, even the largest diffusion only reach a small fraction of the potential audience, e.g., even the largest Twitter diffusion only reaches a small fraction of users \cite{gleeson_branching_2021}.

We consider an undirected network with two communities; however, our approach can be easily applied to an arbitrary number of communities.
The networks with communities should satisfy the following conditions: (a) they can be well described via a configuration model; (b) they have locally tree-like structure (either no or a limited amount number of triangles, clustering, in the network). 

We describe the networks by the degree distributions within each community and between communities. Let $D^{(1)}_1$ be the random variable for the number of connections a randomly chosen node from community 1 has to other nodes within community 1, i.e.,  let $P(\dis{D_{1}}{1} = k)$ be the probability that a random node from community 1 has $k$ connections to nodes in community 1. This random variable will follow the \textit{in-community degree distribution} for community 1. Now let $ D^{(1)}_2 $ be the random variable for the number of connections a randomly chosen node from community 1 has to nodes in community 2, which follows the \textit{between-community degree distribution}. In this notation, the superscript indicates the community of the node, while the subscript denotes the community of the nodes to which the end of the edges terminates. Such notation will be used consistently throughout the paper. Likewise, the in-community and between-community degree of a node for community 2 will be denoted by the random variables $D^{(2)}_2$ and $D^{(2)}_1$, respectively. The connection process follows a configuration model \cite{newman_networks_2018}, where each node has a random number of edges drawn independently from the nodes in- and between-community degree distributions. For example, a node in community 1 has the following probability of having $k$ in-community edges and $j$ between-community edges $P[D^{(1)}_1=k] \times P[D^{(1)}_2=j]$, as these distributions are treated as being independent of each other. The end of these $k$ and $j$ edges are randomly connected to $k$ other nodes inside community 1 and $j$ nodes in community 2.

Our goal is to calculate typical statistical characteristics of the diffusion process, such as total cascade size distribution (also referred to as the total outbreak size or total progeny distribution) and extinction probability. Additionally, we aim to extract quantities that are particularly interesting for diffusion on community-based networks, such as the probability of a contagion ``escaping'' its seed community or the probability of reintroduction of a contagion into a community where it has previously stopped spreading.

 To calculate these quantities, we will describe the diffusion process through a multi-type branching process. \textit{A multi-type branching process (MTBP)} is a discrete Markov process that tracks the active nodes from each community at each time step as random variables \cite{caswell_matrix_2002,Keating_Gleeson_OSullivan_2023,kimmel_branching_2002}. We will refer to the active nodes in community 1 as \textit{``type 1'' active nodes}, and the active nodes in community 2 will be called \textit{``type 2'' active nodes}. At each time step, also referred to as a \textit{generation} of the process, each type of node independently produces a random number of active \textit{children} node. These active children are refereed to as the active parent's offspring in the Branching Processes literature \cite{caswell_matrix_2002}. The offspring follow a specific probability distribution, known as the \textit{offspring distribution}. These offspring then continue the process, generating new descendants in subsequent generations, as illustrated in \mbox{Fig.\ \ref{fig:model}b)}. The primary components in this formulation are the offspring distributions, one specified for each type of node that we track. These distributions are then used to determine the number of active nodes of each type in a given generation (time step). The goal, generally in a Branching Processes, is to calculate the probability distribution of the total number of offspring of each type produced in each generation, as this fully determines the dynamics of the branching process \cite{caswell_matrix_2002}. Once this distribution is obtained, we can proceed with calculating key characteristics, such as cascade size, extinction probabilities, etc.

In the following sections, we develop a framework based on multi-type branching processes to describe cascade dynamics in each community separately as well as in the entire network. We first develop our method for a simple Stochastic Block Model (SBM) \cite{holland_stochastic_1983}. The model has the attractive property that for a sufficiently large network, the degree distributions inside the communities and between the communities can be well approximated by a Poisson distribution. Later, we explain how to extend the approaches to a custom large graph with known in-community and between-community degree distributions.

\section{Probability-generating function framework in a Poisson-distributed network}\label{sec:pgf_framework}

First, let us consider a network consisting of two communities that are Erd\H{o}s--R\'enyi random graphs. For simplicity, we assume that both communities have $\lambda_{\text{in}}$ expected number of internal edges. The edges between communities are also Poisson distributed with  $\lambda_{\text{out}}$ expected number of edges, where $\lambda_{\text{in}} > \lambda_{\text{out}}$. Meaning that, for example, a nodes in community 1 in-community degree is given by the random variable  $\dis{D_1}{1} \sim Poi(\lambda_{in})$ and a nodes between-community degree is given by the random variable $\dis{D_2}{1}\sim Poi(\lambda_{out}).$ This corresponds to a \textit{Stochastic Block Model} (SBM) \cite{holland_stochastic_1983}.

A convenient way to describe the offspring distribution is through the use of \textit{probability-generating functions} (PGFs), which maps all the probabilities of a discrete random variable onto a single power series.
The goal of this section is to construct a PGF for the number of nodes active in each community at any time. 
We let $N_1(t)$ be random variable for the number of active nodes in community 1 at generation $t$ and $N_2(t)$ be the random variable number of active nodes in community 2 at generation $t$ (see Fig.~\ref{fig:model}b). We then introduce the probability generation function $\G{N}{t}$ for $\boldsymbol{N}(t)=(N_1(t), N_2(t))$ that keeps track of the joint probabilities $P[N_1(t)=n,N_2(t)=m]$ as
    \begin{equation}
        \G{N}{t}(s_1,s_2) = \sum_{n=0}^\infty \sum_{m=0}^\infty P[N_1(t)=n,N_2(t)=m] s_1^n s_2^m,
        \label{eq:PGF_def}
    \end{equation}
where $s_1$ and $s_2$ are dummy variables in the power series for type 1 and type 2 offspring, respectively \cite{caswell_matrix_2002,keating_multitype_2022}. Note that all PGFs will be labeled $G_{\cdot}(\vec{s})$, which track the probabilities for the random variable $\cdot$ with dummy variables $\vec{s}$.

To construct $\G{N}{t}$, we begin by constructing the PGF for the offspring distribution generated by a single individual within the network. Let the random variable $X^{(1)}_1$ denote the number of offspring of type 1 produced in the next generation by a single active individual of type 1, and $X^{(1)}_2$ the number of offspring of type 2 produced by a node of type 1. To clarify the notation, we note that the superscript identifies the type of the parent node (what community the parent is from) while the subscript points to the type of offspring produced (what community the offspring are in) --- which is the same name convention used for the degree distributions. Then the PGF generating the offspring distribution produced by a single node of the first type, $G_{\boldsymbol{X^{(1)}}}$, is given by the bi-variate power series

 \begin{equation}
        G_{\boldsymbol{X^{(1)}}}(s_1,s_2) = \sum_{n=0}^\infty \sum_{m=0}^\infty P[X^{(1)}_1=n,X^{(1)}_2=m] s_1^n s_2^m,
        \label{eq:G_X1}
\end{equation}
where $\boldsymbol{X^{(1)}}=(X^{(1)}_1, X^{(1)}_2)$. Similarly, the offspring distribution produced by a single node of type 2 is generated by
    \begin{equation}
        G_{\boldsymbol{X^{(2)}}}(s_1,s_2) = \sum_{n=0}^\infty \sum_{m=0}^\infty P[X^{(2)}_1=n,X^{(2)}_2=m] s_1^n s_2^m.
        \label{eq:G_X2}
    \end{equation}
The PGFs $G_{\boldsymbol{X^{(1)}}}$ and $G_{\boldsymbol{X^{(2)}}}$ are easy to calculate if the in-community and between-community degree distributions of the networks are known. Moreover, if the number of offspring from types 1 and 2 are independent of each other, as is the case in the SBM, the PGF will factor into the product of two PGFs as follows
\begin{equation}
 \begin{aligned}
        G_{\boldsymbol{X^{(1)}}}(&s_1,s_2) 
        = G_{X_1^{(1)}}(s_1)G_{X_2^{(1)}}(s_2) \\
        &= \sum_{n=0}^\infty P[X^{(1)}_1=n] s_1^n  \sum_{m=0}^\infty P[X^{(1)}_2=m]s_2^m.
        \label{eq:G_X1_v2}
\end{aligned}    
\end{equation}
In the SBM case, it is easy to obtain a closed-form expression for $\g{X}{1}$ and $\g{X}{2}$. 
As the size of the network tends to infinity for an SBM, both in-community and between-community degree distributions will have a Poisson distribution with probabilities 

\begin{equation}
    \begin{aligned}
    P[\dis{D_{1}}{1} = k] &= \dfrac{\lambda_{in}^k e^{-\lambda_{in}}}{k!}, \quad \text{and} \\
    P[\dis{D_{2}}{1} = k] &= \dfrac{\lambda_{out}^k e^{-\lambda_{out}}}{k!}.
    \label{eq:3_Poisson_dist}
\end{aligned}
\end{equation}
Now let us calculate the offspring distribution produced by a node of community 1, but note the same logic will apply to the nodes of the second community. First, we calculate the number of activation a node of community 1 produces inside its community. The probability of a node having $k$ neighbors inside community 1 is given by Eqn. \eqref{eq:3_Poisson_dist}. Each of these neighbor nodes is activated independently with probability $\rho$, meaning that the probability of having $x$ successfully activated nodes in community 1 from an active node with in-degree $k$ is $\dfrac{\lambda_{in}^k e^{\lambda_{in}}}{k!}{k \choose x}(1-\rho)^{k-x}(\rho)^{x}$. Summing over all possible values of $k$ yields the probability of having $x$ activated offspring nodes inside community 1 as follows
\begin{equation}
    P[\dis{X_{1}}{1} = x] =  \dfrac{(\rho\lambda_{in})^x e^{-\rho\lambda_{in}}}{x!},
\end{equation}
which itself is Poisson distributed. Following the same argumentation, the probability of activation in community 2 produced by a node of community 1 is $ P[\dis{X_{2}}{1} = y] =  \dfrac{(\rho\lambda_{out})^y e^{-\rho\lambda_{out}}}{y!}$. 
The PGF generating offspring distribution produced by a node of community 1 is then

\begin{equation}
\begin{aligned}
    &\g{X}{1}(s_1, s_2) \\
    &= \sum_{n=0}^{\infty} \frac{(\rho\lambda_{in})^n e^{-\rho\lambda_{in}}}{n!}s_1^n \sum_{m=0}^{\infty} \frac{(\rho\lambda_{out})^m e^{-\rho\lambda_{out}}}{m!}s_2^m \\
    &= e^{\rho\lambda_{\text{in}}(s_1-1)}e^{\rho\lambda_{\text{out}}(s_2-1)}.
    \label{eq:g_X1_intro}
\end{aligned}
\end{equation}
Similarly, the PGF for the offspring distribution produced by an individual of type two is 

\begin{equation}
    \g{X}{2}(s_1, s_2) =e^{\rho\lambda_{\text{out}}(s_1-1)}e^{\rho\lambda_{\text{in}}(s_2-1)}.
    \label{eq:g_X2_intro}
\end{equation}
The formulas \eqref{eq:g_X1_intro} and \eqref{eq:g_X2_intro} give us expressions for the PGFs generating the offspring distribution of each type of node for an SBM, and these are the main building blocks of our method. Once we derive them, calculating $\G{N}{t}$ becomes easy.

It is important to mention that in the general case of a non-Poissonian network, we will have to deal with a larger number of types in the MTBP because the seed node and the nodes activated in the subsequent generations will have different offspring distributions. Moreover the offspring distributions of nodes in the subsequent generations will depend on the way the node was activated and we will need to account for the excess degree distributions when deriving the PGF \cite{newman_networks_2018}. The reason we start with a Stochastic Block Model is due to a convenient property of Poisson networks: their excess degree also follows a Poisson distribution.\footnote{The degree distribution of a Poissonian network is $D(k)=e^{-\langle k \rangle} \langle k \rangle^k/k!$, where $\langle k \rangle$ is the mean-degree. The excess degree distribution, $\widetilde{D}(k)$, by definition is given by $\widetilde{D}(k)=D(k+1)(k+1)/\langle k \rangle$. The excess degree distribution for the Poissonian network is then $\widetilde{D}(k)= e^{-\langle k \rangle}\langle k \rangle^k/k!$, which is exactly the degree distribution $D(k)$ \cite{newman_networks_2018}.} This eliminates the necessity of accounting for the excess degree distribution and makes our final formulas more compact. However, we will have to return to the discussion of the excess degree distributions in Sec.~\ref{sec:Custom_network} where we extend our analysis to general networks. 

After deriving $\g{X}{1}$ and $\g{X}{2}$ we can proceed with calculating the distribution of all active nodes at any given time step $t$ as follows. The probability generating function $\G{N}{t}$ is calculated from the following function iteration:
    \begin{equation}
        \G{N}{t}(s_1, s_2) = \G{N}{t-1}\left(G_{\boldsymbol{X^{(1)}}}(s_1,s_2),G_{\boldsymbol{X^{(2)}}}(s_1,s_2)\right).
        \label{eq:G_N_calculate}
    \end{equation}

We have the probability generating function for the number of active nodes of type 1 and type 2 at time $t-1$. Using this, we wish to see how many offspring we will have in the following generation using $\bs{\dis{X}{1}}$ and $\bs{\dis{X}{2}}$. This characterization of the branching process is called the forward approach in Ref.~\cite{kimmel_branching_2002}, in analogy with the forward Chapman-Kolmogorov equation of Markov processes. It is the common way in which one can derive the number of active nodes in a generation once we have obtained the offspring distributions for each type in MTBP. Refer to Caswell \cite{caswell_matrix_2002} for a derivation of this relationship. We set the initial condition $\G{N}{0}(s_1, s_2) = (s_1)^1(s_2)^0$ corresponding to a single active individual in community 1 and none in community 2.

The utility of this PGFs formulation is that the PGF $\G{N}{t}$ contains information about all the probabilities of the number of active nodes in each community at any generation $t$, and these probabilities can be easily recovered from the PGF. By the definition, the joint probability $P[N_1(t) = n, N_2(t) = m]$ is the corresponding derivative of $\G{N}{t}$, as follows:
\begin{widetext}
    \begin{equation}
        P[N_1(t) = n, N_2(t) = m] = \frac{1}{n! m!}\left[ \frac{\partial^{n}}{\partial s_1^{n}}\frac{\partial^{m}}{\partial s_2^{m}} \left( \G{N}{t}(s_1, s_2) \right) \right]_{(s_1 = 0, s_2=0)}.
        \label{eq:3_probability_definition}
    \end{equation}
\end{widetext}
However, numerical differentiation of PGFs can prove to be unstable, especially for higher-order derivatives. Thus, in practice we use the discrete Fourier transform to evaluate probabilities for the first $K$ values for $N_1(t)$ and $N_2(t)$ as follows \cite{Keating_Gleeson_OSullivan_2023, Gleeson_Ward_OSullivan_Lee_2014, Cavers_1978}:
\begin{widetext}
\begin{equation}
    P[N_1(t) = n, N_2(t) = m] \approx \frac{1}{K^2} \sum_{k_1=0}^{K-1} \sum_{k_2=0}^{K-1} \G{N}{t}(e^{2\pi i k_1/K}, e^{2\pi i k_2/K}) e^{2\pi i n k_1/K}e^{2\pi i m k_2/K}.
    \label{eq:3_probability_FFT}
\end{equation}
\end{widetext}
How to make the transition from Eqn. \eqref{eq:3_probability_definition} to \eqref{eq:3_probability_FFT} is explained in App.~\ref{sec:FFT}.

To obtain the probability distribution of the number of active nodes in each of the communities separately, $P[N_1(t) = n]$ and $P[N_2(t) = m]$, we simply marginalize the probabilities in the community that is not of interest. Thus, the probability $P[N_1(t) = n]$ will be recovered from $\G{N}{t}(s_1,1)$, while the probability $P[N_2(t) = m]$ will be recovered from $\G{N}{t}(1,s_2)$, e.g.,
\begin{equation}
    P[N_1(t)=n] \approx \frac{1}{K} \sum_{k=0}^{K-1} G_{N(t)}(e^{2\pi i k/K},1) e^{2\pi i n k/K}.
\end{equation}

Moreover, to obtain the probability of the total number of active nodes,  $P[N_1(t)+N_2(t) = n]$, we make the substitution $s_1=s,\; s_2=s$ and apply the discrete Fourier transform to $\G{N}{t}(s,s)$. For a more detailed explanation of how to recover the probabilities from $\G{N}{t}$ see App.~\ref{sec:appendix_recovering_probabilities}. 

In summary, using the recursive Eqn.~\eqref{eq:G_N_calculate}, together with PGFs for the offspring distributions \eqref{eq:g_X1_intro} and \eqref{eq:g_X2_intro} we can determine $\G{N}{t}$ for any given generation. Using this we can then obtain some quantities of interest like the extinction probabilities, hazard function and the introduction probabilities (which is the focus on the following sections). Additionally, once we have $\G{N}{t}$, we compute the actual probabilities with the use of discrete Fourier transform \eqref{eq:3_probability_FFT}. Once we have the probability distributions for the number of active nodes of each type at any time point, we possess complete knowledge about the process. 
   
\subsection{Extinction probability and hazard function} \label{sec:ext_hazard}

Our PGF approach allows us to straightforwardly calculate common survival analysis probabilities, such as the extinction probability and hazard function. Let us calculate the \textit{extinction probability}, the probability that the process is extinct at generation $t$ in both communities, $q(t)=P[N_1(t)=0,N_2(t)=0]$. From the properties of PGFs, it follows that
\begin{equation}
\begin{aligned}
    q(t)&
    = P[N_1(t)=0,N_2(t)=0] \\&
    =\sum_{n=0}^\infty \sum_{m=0}^\infty P[N_1(t)=n,N_2(t)=m] (0)^n (0)^m \\
    &= \G{N}{t}(0,0).
    \label{eq:q_both}
\end{aligned}    
\end{equation}
We can also straightforwardly evaluate the probability of the process being extinct at generation $t$ only in community 1 or 2, respectively as
\begin{equation}
 \begin{aligned}
    q_1(t)&=P[N_1(t)=0] = \G{N}{t}(0,1) \text{ and } \\
    q_2(t)&=P[N_2(t)=0] = \G{N}{t}(1,0).
    \label{eq:q_communities}
\end{aligned}    
\end{equation}
With a bit more effort, we can compute the \textit{hazard function}, which is the probability that the process is extinct in a community at time $t$ given that it is not extinct at $t-1$, while possibly still spreading in the other community. We start with the hazard function for community 1, $h_1(t)=P[N_1(t)=0|N_1(t-1) + N_2(t-1)>0]$. This can be expressed as
\begin{equation}
\begin{aligned}
h_1(t) 
&= 1 -\sum_{k=1}^\infty P[N_1(t)=k|N_1(t-1) + N_2(t-1)>0] \\
&= 1 -\dfrac{\sum_{k=1}^\infty P[N_1(t)=k]s^k}{P[N_1(t-1)+N_2(t-1)>0]}.
\end{aligned}
\end{equation}
Noting that $\sum_{k=1}^\infty P[N_1(t)=k] = 1- \G{N}{t}(0,1)$ and $P[N_1(t-1) + N_2(t-1) =0] = \G{N}{t-1}(0,0)$ we obtain
\begin{equation}
    h_1(t) = \frac{\G{N}{t}(0,1)-\G{N}{t-1}(0,0)}{1-\G{N}{t-1}(0,0)},
\end{equation}
which is the hazard function for community 1. In the same way, the hazard function for the second community is
\begin{equation}
    h_2(t) = \frac{\G{N}{t}(1,0)-\G{N}{t-1}(0,0)}{1-\G{N}{t-1}(0,0)},
\end{equation}
while the hazard function for the entire network is 
\begin{equation}
    h(t) = \frac{\G{N}{t}(0,0)-\G{N}{t-1}(0,0)}{1-\G{N}{t-1}(0,0)}.
\end{equation}

\begin{figure*}[t]
  \centering
  \includegraphics[width=0.9\textwidth]{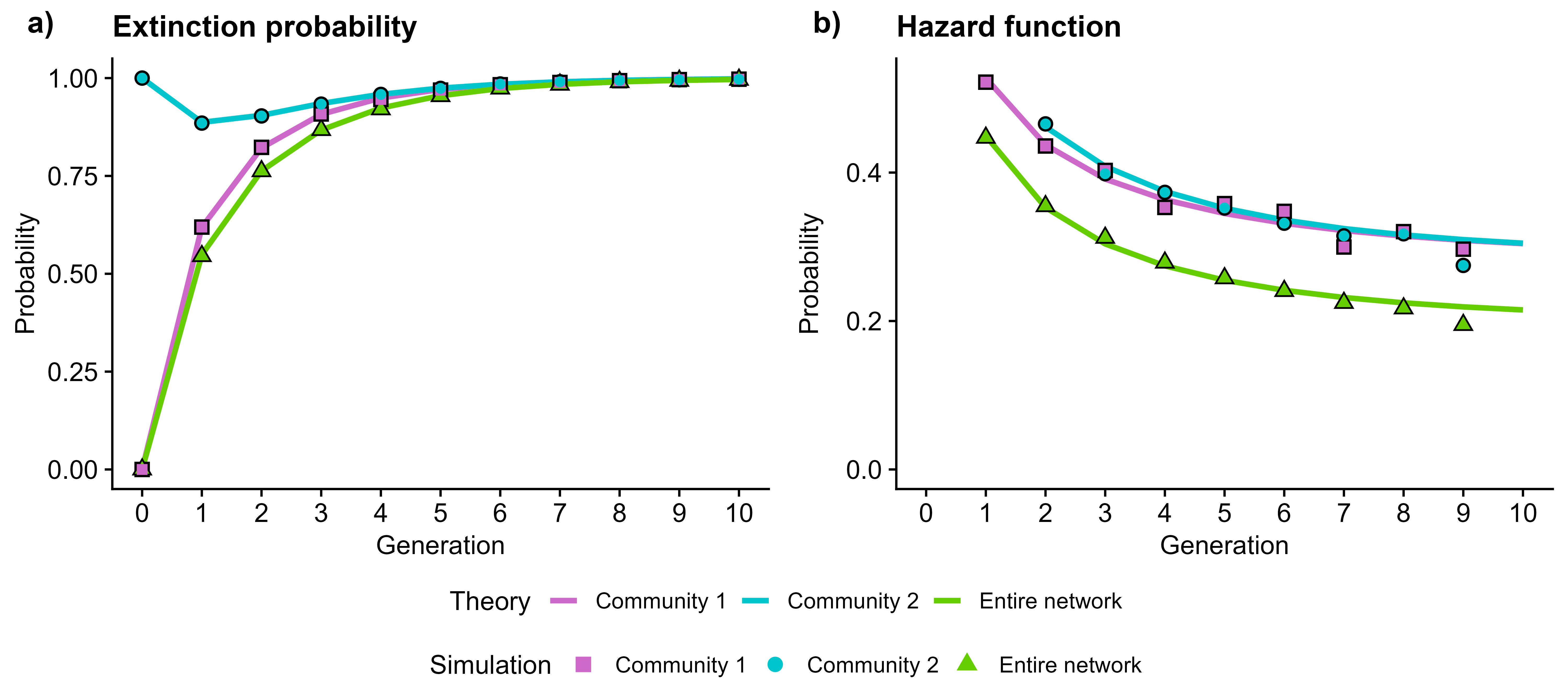}
  \caption{a) Extinction probability for Stochastic Block Model; b) Hazard function for Stochastic Block Model. Here, $\lambda_{\text{in}}=8, \; \lambda_{\text{out}}=2$ and $\rho=0.06$. For numerical simulations, we use branching process simulations with averaging across $5\times 10^{4}$ simulations. The hazard function starts in generation 1 (not 0) as the probability of survival until generation 0 is not defined.}
  \label{fig:q_and_h}
\end{figure*}

In Fig. \ref{fig:q_and_h}a) we show the extinction probability distribution for the entire network and the distributions for each community calculated using Eqns. \eqref{eq:q_both} and \eqref{eq:q_communities}. It compares the analytical results with branching process simulations where $\lambda_{in} =8$, $\lambda_{out}=2$ and $\rho = 0.06$. The hazard functions for the SBM and each community are shown in Fig. \ref{fig:q_and_h}b) and we note the excellent agreement between our simulation and theory curves. For our numerical simulations, we use branching process simulations repeated $5 \times 10^4$ times and we compute all distributions as averages across these simulations. In App.~\ref{sec:Simulations_BP_Network} we demonstrate that the branching process simulation agrees with network-based simulations performed on actual networks with a sufficiently large number of nodes. With both the extinction probabilities and hazard function to hand, we next show that it is possible to construct a community-specific hazard function. This will be an important building block in constructing the probability of reintroducing activations into a community where it had no active nodes in the previous generation. 

\subsubsection{Community specific hazard function}

We can now calculate what we call the \textit{community specific hazard function}, $\widetilde{h}_1(t)$ and $\widetilde{h}_2(t)$. They are hazard functions specified for each community on non-extinction in community 1 or community 2 in the previous time step, i.e., $\widetilde{h}_1(t)=P[N_1(t)=0|N_1(t-1)>0]$ and $\widetilde{h}_2(t)=P[N_2(t)=0|N_2(t-1)>0]$. In contrast, $h_1(t)$ in the last section was the hazard function for community 1 conditioned on the joint non-extinction of the process in the previous time step (i.e., $N_1(t-1)=0$ and $N_2(t-1)$). This can be useful if we care more about when the process stops in a specific community disregarding whether it is still going on in the other one. To derive the expression for $\widetilde{h}_1(t)$ we introduce the random variable $Y_1(t)$ such that $P[Y_1(t)=k] = P[N_1(t)=k|N_1(t-1)>0]$. 
Then $G_{Y_1(t)}$ is the PGF for the number of active nodes at time $t$, conditional on the process being non-zero in community 1, and it is easily calculated from $\G{N}{t}$, as follows
\begin{equation}
    G_{Y_1(t)}(s_{1},s_{2})=\frac{\G{N}{t}(s_{1},s_{2})-\G{N}{t}(0,s_{2})}{1-\G{N}{t}(0,1)}.
\end{equation}
To see this, write out the power series of $\G{N}{t}(s_{1},s_{2})$ and subtract the terms corresponding to $\G{N}{t}(0,s_{2})$, which then is normalized. We can now note a connection between our desired probability $\widetilde{h}_1(t)$ and $G_{Y_1(t-1)}$. The function $G_{Y_1(t-1)}$ tracks the probabilities for the number of active nodes at time $t-1$ given the process is not extinct in community 1. If we find the number of offspring of the active nodes from the $t-1$ generation, we can arrive at the PGF for number of active nodes in generation $t$ conditioned on the process not being extinct in generation $t-1$. This is given by: 
\begin{equation}
G_{Y_1(t-1)}(\g{X}{1}(s_1,s_2), \g{X}{2}(s_1,s_2)).\label{eq:G_H(2)}
\end{equation}
This gives us a way to express the community-specific hazard probability for community 1. Then by calculating the probability of extinction in community 1 (setting $s_1 = 0$) and marginalizing over community 2 (by setting $s_2 = 1$) in generation $t$, we obtain $\widetilde{h}_1(t)$, as required: 
\begin{equation}
\begin{aligned}
\widetilde{h}_1(t)
&= G_{Y_1(t-1)}(\g{X}{1}(0,1),\g{X}{2}(0,1))\\ 
&= \frac{\G{N}{t}(0,1)-\G{N}{t-1}(0,\g{X}{2}(0,1))}{1-\G{N}{t-1}(0,1)}. \label{eq:tilde_h1}
\end{aligned}
\end{equation}
Similarly, the community-specific hazard probabilities for community 2, \(\widetilde{h}_2(t)\), can be derived through an analogous process. It is important to note that we can arrive at all these expressions via simple function iteration followed by the appropriate substitutions of zeros and ones into the dummy variables $s_1$ and $s_2$. Equation \eqref{eq:tilde_h1} allows us to finally arrive at an important quantity of interest, the probability of contagion spreading between communities, which we derive in the next section. 
    
\subsection{Probability of contagion travel between communities}  \label{sec:introduction_p}

When studying the spread of information or disease on a network with communities, a crucial question arises: What is the probability of infection spreading to another community? Here we show how to calculate this probability with our framework. Let $r_1(t)=P[N_1(t)>0|N_1(t-1)=0]$ be the probability of infection in the first community at time $t$, given no infected nodes in that community at time $t-1$, which we will call the \textit{introduction probability} for community 1. Care must be taken in noting what $r_1(t)$ is; it measures the probability of seeing an activated node in community 1, maybe for the first time, but also for the second time, third time, etc. All we know is that there was no infected (active) node in the community in the previous time step. It can be expressed as
\begin{equation}
\begin{aligned}
r_1(t) 
&= 1-P[N_1(t)=0|N_1(t-1)=0] \\
&=1-\phi_1(t), \label{eq:r1}
\end{aligned}
\end{equation}
where we will call $\phi_1(t) = P[N_1(t)=0|N_1(t-1)=0]$ the probability of continued extinction for community 1. Deriving an expression for $\phi_1(t)$ is straightforward by first expanding $q_1(t)$, where we condition on extinction and non-extinction in the previous time step in community 1. Once this is done we can then match terms to quantities we have already derived in previous sections:
\begin{widetext}
\begin{align}
q_1(t) 
    &= \underbrace{P\left[N_1(t)=0|N_1(t-1)=0\right]}_{\phi_1(t)}
    \times  
    \underbrace{P\left[N_1(t-1)=0\right]}_{q_1(t-1)} 
    +
    \underbrace{P\left[N_1(t)=0|N_1(t-1)>0\right]}_{\widetilde{h}_1(t)} 
    \times  
    \underbrace{P\left[N_1(t-1)>0\right]}_{1- q_1(t-1)}  
    \nonumber \\ 
    &= \phi_1(t)q_1(t-1) + \widetilde{h}_1(t)\left[1- q_1(t-1)\right].\label{eq:m_1}
\end{align}
\end{widetext}
Then Eqn. \eqref{eq:m_1} can easily be rearranged to isolate $\phi_1(t)$ yielding
\begin{equation}
\phi_1(t)=\frac{q_1(t)-\widetilde{h}_1(t)\left[1-q_1(t-1)\right]}{q_1(t-1)},\label{eq:c1}
\end{equation}
where $\widetilde{h}_1(t)$ and $q_1(t)$ are calculated by Eqn. \eqref{eq:tilde_h1} and Eqn. \eqref{eq:q_communities}. Using Eqn. \eqref{eq:r1} we can derive the reintroduction probability for community 1 as $r_1(t) = 1 - \phi_1(t)$. The same logic holds when deriving expressions for $\phi_2(t)$ and $r_2(t)$ as we have used above.

\begin{figure*}
  \centering
  \includegraphics[width=0.9\textwidth]{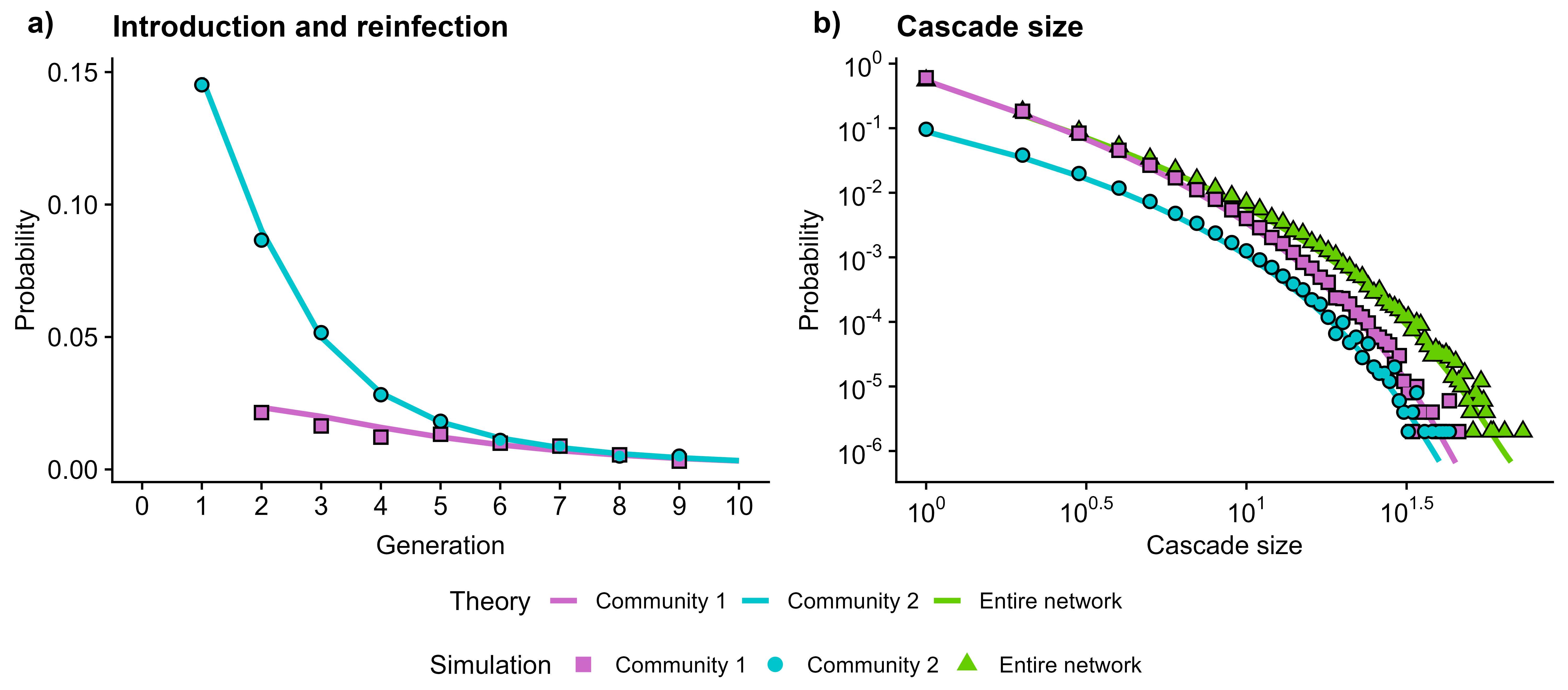}
  \caption{a) Probability of infection spreading to community 2 and the probability of reintroduction back to community 1 for SBM. For numerical simulations, we use branching process simulations where we average across $5\times 10^{5}$ simulations. Here, $\lambda_{\text{in}}=8, \; \lambda_{\text{out}}=2$ and $\rho=0.08$; b) Cascade size distribution for SBM. Here, $\lambda_{\text{in}}=8, \; \lambda_{\text{out}}=2$ and $\rho=0.06$.}
  \label{fig:cascades}
\end{figure*}

We plot the probabilities of introduction and reintroduction for any time point in Fig. \ref{fig:cascades}a) and we observe an excellent agreement with the simulated data. Note that the probability of reintroduction is not defined for the first generation as the process is still alive at the previous generation, i.e., generation zero. 

\subsection{Cascade size distribution}
    
In this section, we calculate the overall size of outbreaks, i.e., the cascade size distribution, with our framework. The cascade size distribution is also referred to as the total progeny distribution of a seed within the literature on branching processes \cite{kimmel_branching_2002}. This cascade size distribution can simply be found by introducing a counter type into our multi-type branching process that tracks the total number of active nodes that have been created for each community. The authors found the idea of using a counter in this manner in Ref. \cite{haccou2005branching}, where it was introduced as a means of calculating the expected population size in a MTBP. We adapt it here for cascade distribution calculations.

To track the cascade size distribution we introduce $C_1$, a random variable tracking the total number of active nodes in a cascade from community one, such that $P[C_1(t)=k] = P[\sum_i^t N_1(i) = k]$ and, similarly, $C_2(t)$ is defined similarly as $P[C_2(t)=k] = P[\sum_i^t N_2(i) = k]$. Both $C_1(t)$ and $C_2(t)$ are tracked using new counter-types mapped to the dummy variables $c_1$ and $c_2$ within a PGF. These counter-type function as follows: every time a node becomes infected in community 1, a counter type, $c_1$, is created. Similarly, for every newly infected node in community 2, a counter type, $c_2$, is created. All counter-types are then forced to persist between generations. This translates to a probability generation function that maps the joint probability distribution of variables $N_1, N_2, C_1$ and $C_2$ which is given by
\begin{widetext}
\begin{align}
        \G{C}{t}(s_1,s_2, c_1, c_2) = \sum_{i,j,k,l} 
        P[N_1(t)=i,N_2(t)=j,C_1(t)=k, C_2(t)=l] s_1^i s_2^j c_1^k c_2^l.
        \label{eq:G_C_def}
\end{align}
\end{widetext}
The PGF $\G{C}{t}$ is effectively an extension of $G_{N(t)}$, which includes the two new counter types that track the number of active nodes in each community. Just as we had a recursive relationship $G_{N(t)}$ given by Eqn. \eqref{eq:G_N_calculate}, we have a recursive relationship for $\G{C}{t}$ given by
\begin{widetext}
\begin{equation}
    \G{C}{t}(s_1,s_2,c_1,c_2) =
    \G{C}{t-1}( 
    \g{X}{1}(
    \overbrace{\rule{0pt}{1.5ex}{s_1 c_1}}^{\mathclap{\text{When a type-1 node is created a type-1 counter is created also.}}},
    s_2 c_2), \g{X}{2}(s_1 c_1,s_2 c_2),
    \underbrace{c_1, c_2}_{\mathclap{\text{Number of counters persist from $t-1$ to $t$.}}}
    )
\end{equation}
\end{widetext}
Note that each counter type $s_1$ and $s_2$ always produce exactly one copy of themselves between generations, thus accounting for the total cascade size. 
The initial condition is then given as $\G{C}{0}(s_1, s_2,c_1, c_2) = s_1 c_1$, where we have one active node in community one ($s_1$) and a counter for that node ($c_1$).

As we are interested in the total cascade size, we substitute $c_1 = c$ and $c_2= c$ in $\G{C}{t}(1, 1,c, c)$. We then marginalize $s_1$ and $s_2$ in Eqn.~\eqref{eq:G_C_def} to calculate the actual probabilities $P[C_1(t) + C_2(t) = n]$, which is a univariate PGF. The probability distribution can then be estimated from the univariate PGF using the inverse fast Fourier transform to obtain the cascade size distribution as discussed in App. \ref{sec:appendix_recovering_probabilities}.
Moreover, in the same manner, functions $G_{C(t)}(1,1,c,1)$ and $G_{C(t)}(1,1,1,c)$ can yield the probability distribution for cascade sizes in community one and community two, respectively.
As Fig.~\ref{fig:cascades}b) illustrates, our framework accurately estimates the expected cascade sizes not only for the entire network but also for each community individually.

\section{Extending the pgf framework to a general network with communities}\label{sec:Custom_network}

In this section, we extend the analysis presented in Sec.~\ref{sec:pgf_framework} to a broader class of networks with community structure, where their degree distributions do not necessarily follow the Stochastic Block Model, i.e., the in- and between community degree distributions are not Poisson. While the networks constituting communities can be general, they are required to have a local tree-like structure and be well-described via the configuration model. In principle, it is possible to relax these assumptions, but this is left to future work.

Incorporating networks with a general degree requires careful consideration of the origin of node activation --- whether from within or outside its own community. This in turn, will lead to the use of excess offspring distributions. Below, we explain how to incorporate this into our framework.
As in the previous section, the first step is to derive the offspring distributions for each type of node we track. Let us start by calculating the offspring distribution for the seed node in community 1. For our framework to work we need to know (or be able to extract) the in-community degree of each node, $D^{(1)}_1$ and $D^{(2)}_2$, and the between-community degree of each node, $D^{(1)}_2$ and $ D^{(2)}_1$. We then construct the PGFs corresponding to degree distributions, e.g.,
\begin{equation}
\begin{aligned}
    G_{\dis{D_1}{1}}(s) &= \sum_{k=1}^{\infty}P[\dis{D_1}{1}=k]s^k, \\ \text{and} \quad G_{\dis{D_2}{1}}(s) &= \sum_{k=1}^{\infty}P[\dis{D_2}{1}=k]s^k.    
\end{aligned}    
\end{equation}
When we have this, we proceed with calculating the PGF for the offspring distribution as follows. Under the ICM, the probability that a single node, who is in contact with an active node, will become active themselves is defined by the PGF for a Bernoulli distribution $G_I(s) = 1-\rho + \rho s$, where $I \sim Ber(\rho)$. The number of active individuals in- and between-community one defines randomly stopped sums $\dis{X_1}{1} = \sum_{k=0}^{\dis{D_1}{1}} I$ and $\dis{X_2}{1} = \sum_{k=0}^{\dis{D_1}{2}} I$, respectively. With the PGF for the randomly stopped sum of the offspring distributions can be written compactly as 
\begin{equation}
    \begin{aligned}
    G_{\dis{X_1}{1}}(s) &= G_{\dis{D_1}{1}}(G_I(s)), \quad 
    \\ \text{and}\quad G_{\dis{X_2}{1}}(s) &= G_{\dis{D_2}{1}}(G_I(s)). \label{eq:off_rss}
\end{aligned}
\end{equation}
This gives us, respectively, the probability generating functions for the offspring distribution from a randomly active node in community 1 inside its community and between communities \cite{grimmett2020probability}. 
Similarly, if we start in community 2, the offspring distributions for the seed node will be $G_{\dis{X_2}{2}}(s) = G_{\dis{D_2}{2}}(G_I(s))$ and $G_{\dis{X_1}{2}}(s) = G_{\dis{D_1}{2}}(G_I(s))$. PGFs are particularly useful when dealing with randomly stopped sums, see Ref. \cite{grimmett2020probability} for a brief useful introduction. In fact, the reader might find it interesting to note that Eqns. \eqref{eq:g_X1_intro} and \eqref{eq:G_N_calculate} are both randomly stopped sums of the same form as Eqn. \eqref{eq:off_rss}.

Now, let us consider what happens when the activations spread beyond the seed generation. According to ICM, a node cannot be activated again once it has been activated in a previous time step. If we have an active node, who was activated by a neighbor in the previous generation, the number of inactive (susceptible) nodes that can then go on to be activated differs from the seed generation.
To account for this, we need to use the appropriate excess degree distribution.

\begin{figure*}
    \centering
    \includegraphics[width=1\linewidth]{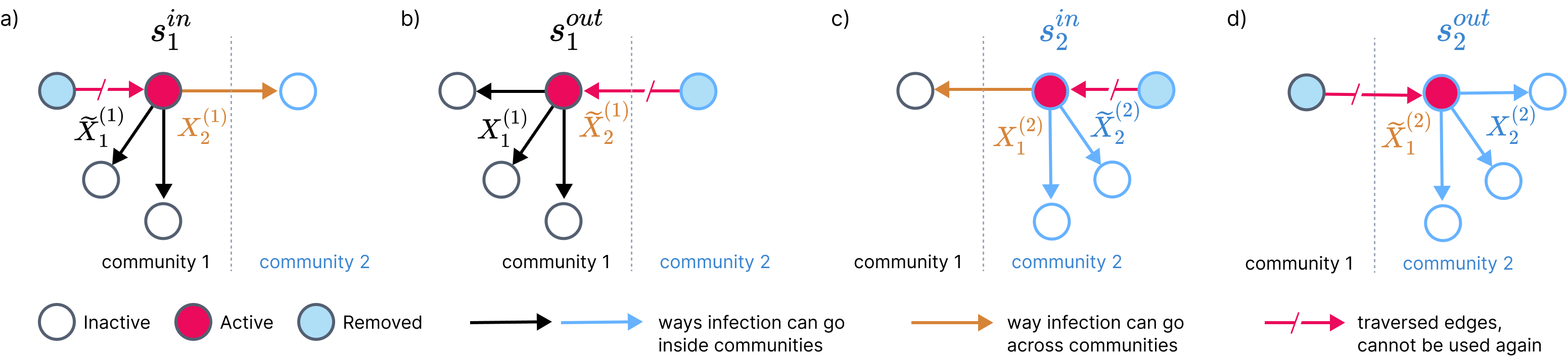}
    \caption{Schematic illustrating four types of offspring tracking in the model: a) $s_1^{in}$ represents offspring in community 1 produced by traversing the edge within community 1. It produces $\widetilde{X}^{(1)}_1$ and $X^{(1)}_2$ offspring in the next time step; b) $s_1^{out}$ is offspring in community 1 produced by traversing the edge from community 2, producing $X^{(1)}_1$ and $\widetilde{X}^{(1)}_2$ offspring in the next time step; c) $s_2^{in}$ is offspring in community 2 produced by traversing the edge within community 2. It creates $X^{(2)}_1$ and $\widetilde{X}^{(2)}_2$ offspring in the next time step; d) $s_2^{out}$ is offspring in community 2 produced by traversing the edge from community 1. It creates $\widetilde{X}^{(2)}_1$ and $X^{(2)}_2$ offspring in the next time step. The red arrow represents the edge through which the node was infected. This edge cannot be used again, so we use the excess degree distribution for the communities with a traversed edge. Black, blue and yellow arrows show the ways that infection can proceed in the next generation.}
    \label{fig:offspring_types}
\end{figure*}

Mathematically, the relationship between the PGF of a distribution and its excess distribution is an elegant and simple one \cite{newman_networks_2018}. If we have a PGF for the offspring distribution, with associated random variable  $\dis{X_1} {1}$ and PGF $G_{\dis{X_1}{1}}(s)$, its excess offspring PGF will be given by 
\begin{equation}
    G_{\dis{\widetilde{X}_1}{1}}(s) = \frac{G^{\prime}_{\dis{X_1}{1}}(s)}{G^{\prime}_{\dis{X_1}{1}}(0)}. \label{eq:pgftoexcesspfg}
\end{equation}
To account for this in our multi-type branching process, we need to also consider where a node activation originated from --- either by another node from their community or from outside their community, as illustrated in Fig.~\ref{fig:offspring_types}. Consider an active node in community 1 after generation $t=0$. This node could have been activated in two ways: 
\begin{enumerate}[(1)]
    \item If it was activated from inside its own community, then to correctly track activations from it, we need to use the excess offspring distribution for community 1,  with associated random variable $\dis{\widetilde{X}_1}{1}$, (as we have used up one possible link that the activations could travel down), and the offspring distribution for between-communities, with associated random variable $\dis{X_2}{1}$, (as we have not used up any edges between communities from which activations could traverse).
    \item If it was activated from outside its own community (by a node in community 2), then to correctly track activations from it, we need to use the offspring distribution for community 1, with associated random variable $\dis{{X}_1}{1}$, (as we have not used up a possible edge that the activations could travel down inside its own community), and the excess offspring distribution for between-communities, with associated random variable $\dis{\widetilde{X}_2}{1}$, (as we have used up an edge between communities from which activations could traverse).
\end{enumerate}
To track each of these cases for community 1 we introduce two types, each representing where a node in community 1 was activated from. The same logic holds for community 2, therefore, we now must keep track of four distinct types of offspring produced, which leads us, as illustrated in Fig.~\ref{fig:offspring_types}, to a 4-type  branching process. The corresponding PGFs will use dummy variables $s_1^{in}$ to track active nodes in community 1 that were activated through a link inside community 1, and $s_1^{out}$ for activations through links from community 2. Likewise, the offspring in community 2 obtained from traveling the edge inside and outside community 2 will be tracked by $s_2^{in}$ and $s_2^{out}$, respectively. Each of these types will have their own offspring distributions and probability-generating functions, which are shown in Tab. \ref{tab:PGFs_explanation}.

The updated notation allows us to specify our multivariate PGFs for the offspring that each type can have. For example, an active node in community 1, where it was activated from within its own community in the previous generation, will use the excess offspring distribution for its own community, 
 with associated random variable $\dis{\widetilde{X}_1}{1}$, and offspring distribution for between the communities, with associated random variable ${X}^{(1)}_{2}$. This leads, naturally, to the PGF for the number of offspring defined by
\begin{align}
    \g{X_{in}}{1}(s_1^{in},s_2^{out}) &=    
       \overbrace{
       G_{\dis{\widetilde{X}_1}{1}}(s_1^{in})
       }^{
       \hspace{-0.2cm}\mathclap{\text{Note use of excess offspring PGF due to }} \atop \mathclap{\text{parent coming from community 1.}}
       }
        \underbrace{
        G_{\dis{X_2}{1}}(s_2^{out})
        }_{
        \mathclap{\text{Offspring PGF due to parent coming }} \atop \mathclap{\text{ from community 1.}}
        }. 
        \label{eq:custom-gx1}
\end{align}
Equation \eqref{eq:custom-gx1} holds as we are treating the degree inside and outside of a community as independent of each other, and the PGF for the sum of independent random variables can be found by taking the product of their PGFs. 

\begin{table*}
\centering
\begin{tabular*}{\textwidth}{@{\extracolsep{\fill}}rl}
\hline
\textbf{PGF} & \textbf{What it generates} \\
\hline
$\g{X}{1}$        & Offspring produced by a seed node. \\
$\g{X_{in}}{1}$   & Offspring produced by a node from community 1 who was activated by traversing an edge inside community 1. \\
$\g{X_{out}}{1}$  & Offspring produced by a node from community 1 who was activated by traversing an edge coming from community 2. \\
$\g{X_{in}}{2}$   & Offspring produced by a node from community 2 who was activated by traversing an edge inside community 2. \\
$\g{X_{out}}{2}$  & Offspring produced by a node from community 2 who was activated by traversing an edge coming from community 1. \\
\hline
\end{tabular*}
\caption{Summary of probability-generating functions generating offspring distributions.}
\label{tab:PGFs_explanation}
\end{table*}

Following the same logic as for the derivation of Eqn.~\eqref{eq:custom-gx1}, we write the expressions for the PGFs for offspring distributions produced by each type of nodes by replacing the degree distributions with their corresponding excess degree distributions where necessary. We obtain

\begin{align}
    \g{X_{out}}{1}(s_1^{in},s_2^{out}) &=   G_{\dis{X_1}{1}}(s_1^{in})              
                                            G_{\dis{\widetilde{X}_2}{1}}(s_2^{out}), \nonumber \\  
                                %
    \g{X_{in}}{2}(s_2^{in},s_1^{out}) &=    G_{\dis{\widetilde{X}_2}{2}}(s_2^{in}) 
                                            G_{\dis{X_1}{2}}(s_1^{out}), 
                                            \nonumber \\
    \g{X_{out}}{2}(s_2^{in},s_1^{out}) &=   G_{\dis{X_2}{2}}(s_2^{in}) 
                                            G_{\dis{\widetilde{X}_1}{2}}(s_1^{out}). 
                                            \nonumber
\end{align}
See Tab. \ref{tab:PGFs_explanation} for the summary of the PGFs for the offspring distributions.

After defining the PGFs for the offspring distributions generated by a single node of each type, we can apply the same approach as in Sec. \ref{sec:pgf_framework} while keeping track of additional random variables. We define the PGF generating the number of active nodes of all types for each generation as
\begin{widetext}
\begin{equation}
     \G{\mathcal{N}}{t}(s_1^{in}, s_1^{out}, s_2^{in}, s_2^{out}) = \sum_{k,l,m,n} P[\mathcal{N}^{in}_1(t)=k, \mathcal{N}^{out}_1(t)=l, \mathcal{N}_2^{in}(t)=m, \mathcal{N}^{out}_2(t)=n] (s_1^{in})^k (s_1^{out})^l (s_2^{in})^m (s_1^{out})^n; \nonumber
\end{equation}
\end{widetext}
here $\boldsymbol{\mathcal{N}}(t) = \left( \mathcal{N}^{in}_1(t), \mathcal{N}^{out}_1(t), \mathcal{N}^{in}_2(t), \mathcal{N}^{out}_2(t)\right)$ is a random variable for numbers of active nodes of each type at generation $t$.
We use $\mathcal{N}(t)$ instead of $N(t)$ to differentiate it from the Poisson networks case. As before, in Sec. \ref{sec:pgf_framework}, the PGF for $\mathcal{N}(t)$ is calculated from the self-contained formula
\begin{widetext}
\begin{equation}
    \G{\mathcal{N}}{t}(s_1^{in}, s_1^{out}, s_2^{in}, s_2^{out}) = \G{\mathcal{N}}{t-1}\left(\g{X_{in}}{1}, \g{X_{out}}{1}, \g{X_{in}}{2}, \g{X_{out}}{2}\right),
\end{equation}
\end{widetext}
We note that if we have a single seed in community 1 then generation 1 will have a slightly different PGF, as no spreading has taken place to require the use of an excess degree distribution. Therefore 
$\G{\mathcal{N}}{1}(s_1^{in}, s_1^{out}, s_2^{in}, s_2^{out}) = 
        G_{\dis{X_1}{1}}(s_1^{in})G_{\dis{X_2}{1}}(s_2^{out}).$

In Sec.~\ref{sec:ext_hazard} and Sec.~\ref{sec:introduction_p}, we derived many characteristic distributions of interest for the Stochastic Block Model. It is important to note that these formulas still hold for $\G{\mathcal{N}}{t}(s_1^{in}, s_1^{out}, s_2^{in}, s_2^{out})$. Moreover, as we are only interested in the distribution of the total number of active nodes in community 1 or community 2, regardless of where they were activated from, we can  easily recover these distributions by noting that setting $s_1^{in}$ and $s_1^{out}$ equal to $s_1$ and $s_2^{in}$ and $s_2^{out}$ equal to $s_2$ reduces $\G{\mathcal{N}}{t}$ to $\G{N}{t}$, i.e.,
\begin{equation}
    \G{\mathcal{N}}{t}(s_1, s_1, s_2, s_2) = \G{N}{t}(s_1, s_2).
\end{equation}
With this formula in mind, we can now apply all the formulas derived for the statistical characteristics for the Stochastic Block Model in Sec.~\ref{sec:pgf_framework} to calculate the same characteristics for a general network of interest. In the following section, we will apply our method to a network with heavy-tailed degree distribution to illustrate our framework for a general network.
    \subsection{Application of pgf framework to a log-normal network with community structure} \label{sec:ln_network_example}

We now focus on a more complex and realistic structure than the simple SBM studied in Sec. \ref{sec:pgf_framework}. We use a network with a discrete log-normal degree distribution as it can exhibit heavier tails than the Poisson distribution, yet it can be easily adjusted to avoid excessively broad distributions. This is important for computational reasons, the broader the degree distribution, the more terms required in the sums defining the PGFs. Additionally, we set the number of between-community edges to be small, with either no connections or only one connection per node between communities. Both communities will be of the same size. This leaves our multi-type branching process description in a general form, allowing for a `plug-and-play' set-up if one wishes to explore other network structures. 

\begin{figure*}[t]
    \centering
    \includegraphics[width=0.9\linewidth]{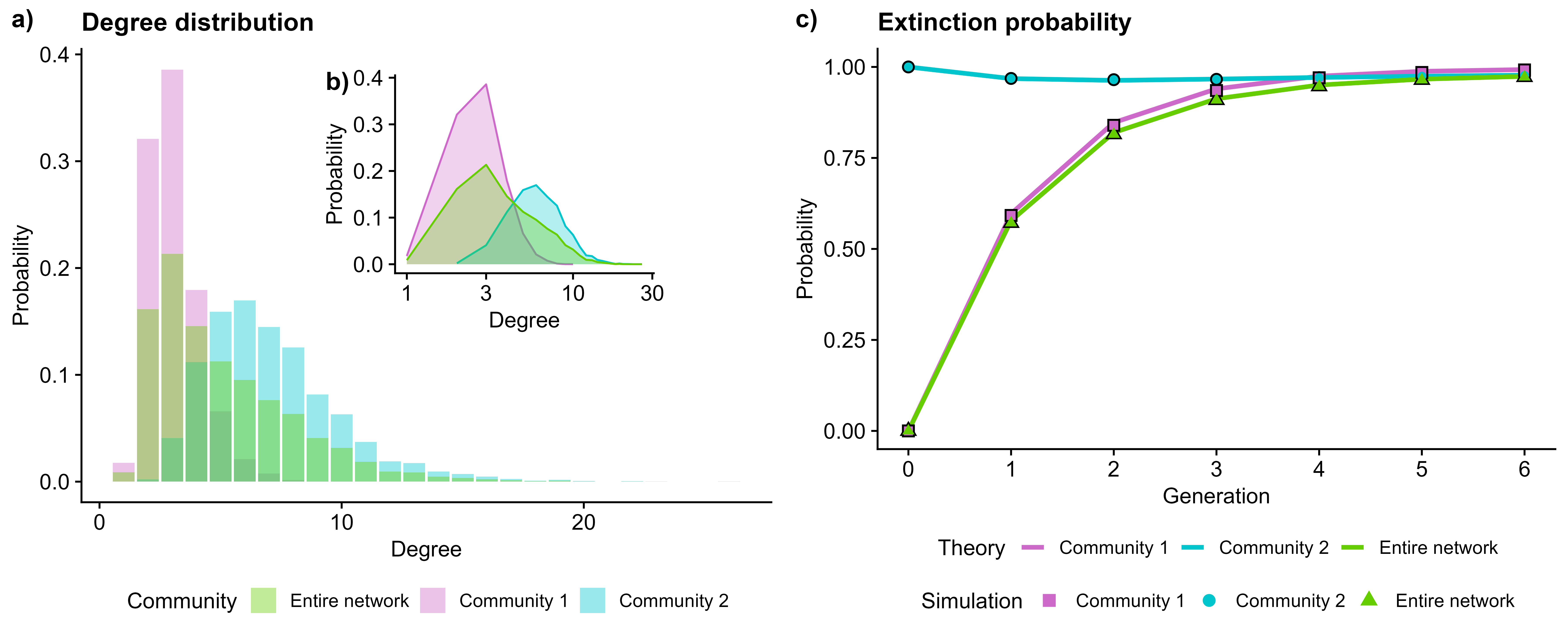}
    \caption{a) Degree distribution for our example log-normal network. Nodes in community 1 have an average degree of $\approx 3$, nodes in community 2 have an average degree of $\approx 7$ and the average degree for a randomly selected node on the network is $\approx 5$. These degree distributions also include the cross community edges; b) Degree distribution for our example log-normal network where the x axis is on a log scale;  c) Extinction probability for log-normal network. Here, $\rho = 0.16$. }
    \label{fig:ln_degree}
\end{figure*}

To construct the required PGF we first need the in- and between-community degree distributions. To extract these, we simulate two networks of $5\,000$ nodes, each with a log-normal distribution internally with $\mu_1 = 0.85$ and $\sigma_1 =0.4$ ($\mu_2 = 1.8$ and $\sigma_2 = 0.4$) for community 1 (and community 2, respectively). As the log-normal is a continuous distribution, the generate numbers are real-valued no integer valued, therefore, we apply the ceiling function to these sampled values. This yields an internal degree structure where the average degree of nodes in community 1 is approximately 3, and the average degree of nodes in community 2 is approximately 7.  Figure \ref{fig:ln_degree}a) shows the degree distribution for our example network. With these distributions, we can formulate the required PGFs $G_{\dis{X_{1}}{1}}$ and $G_{\dis{X_{2}}{2}}$, and then construct the excess offspring distributions $G_{\dis{\widetilde{X}_{1}}{1}}$ and $G_{\dis{\widetilde{X}_{2}}{2}}$.

To generate community structure we add edges between community 1 and community 2 in the following way. We randomly select $n_e$ nodes from community 1 and $n_e$ nodes from community 2, and connect each node across the communities. The value of $n_e$ is set to be $100$. The probability of a node having a cross-community edge is then given by probability $p_e = n_e / 5\,000 = 0.02$. As both communities are of the same size, this yields the same between-community degree distribution, which will have PGF $G_I(s) = 1-p_e + p_e s$, leading to the between-community offspring distributions 
\begin{align}
G_{\dis{X_{2}}{1}}(s) = G_{\dis{X_{1}}{2}}(s) 
            &= 1-p_e + p_e (G_I(s)) \nonumber \\
            &=  1-p_e + p_e (1 - \rho + \rho s). \nonumber    
\end{align}

We can easily show that the excess offspring distributions for between-community spread are given by $G_{\dis{\widetilde{X}_{2}}{1}}(s) = G_{\dis{\widetilde{X}_{1}}{2}}(s) = s^0=1$. This means that, for example, if a node in community 1 was infected by a edge emanating from community 2 it is certain to have no other edges connecting to community 2 (as we have used the only edge to activated it). Now that we have the four probability generation functions, we can iterate them to find the probability generation function for the number of active nodes at time $G_{\mathcal{N}(t)}$. 

Once we constructed $G_{\mathcal{N}(t)}$, we can proceed to calculate characteristics of interest, such as cascade sizes and extinction probabilities. In Fig.~\ref{fig:ln_degree}b) we show the extinction probabilities calculated for the log-normal network using equations \eqref{eq:q_both} and \eqref{eq:q_communities}. We observe excellent agreement with the simulated distributions. In the next section, we focus on the cascade size distribution.

    
\subsection{The community seeding effect on cascades sizes}\label{sec:effect_community}

\begin{figure*}[t]
    \centering
    \includegraphics[width=0.8\linewidth]{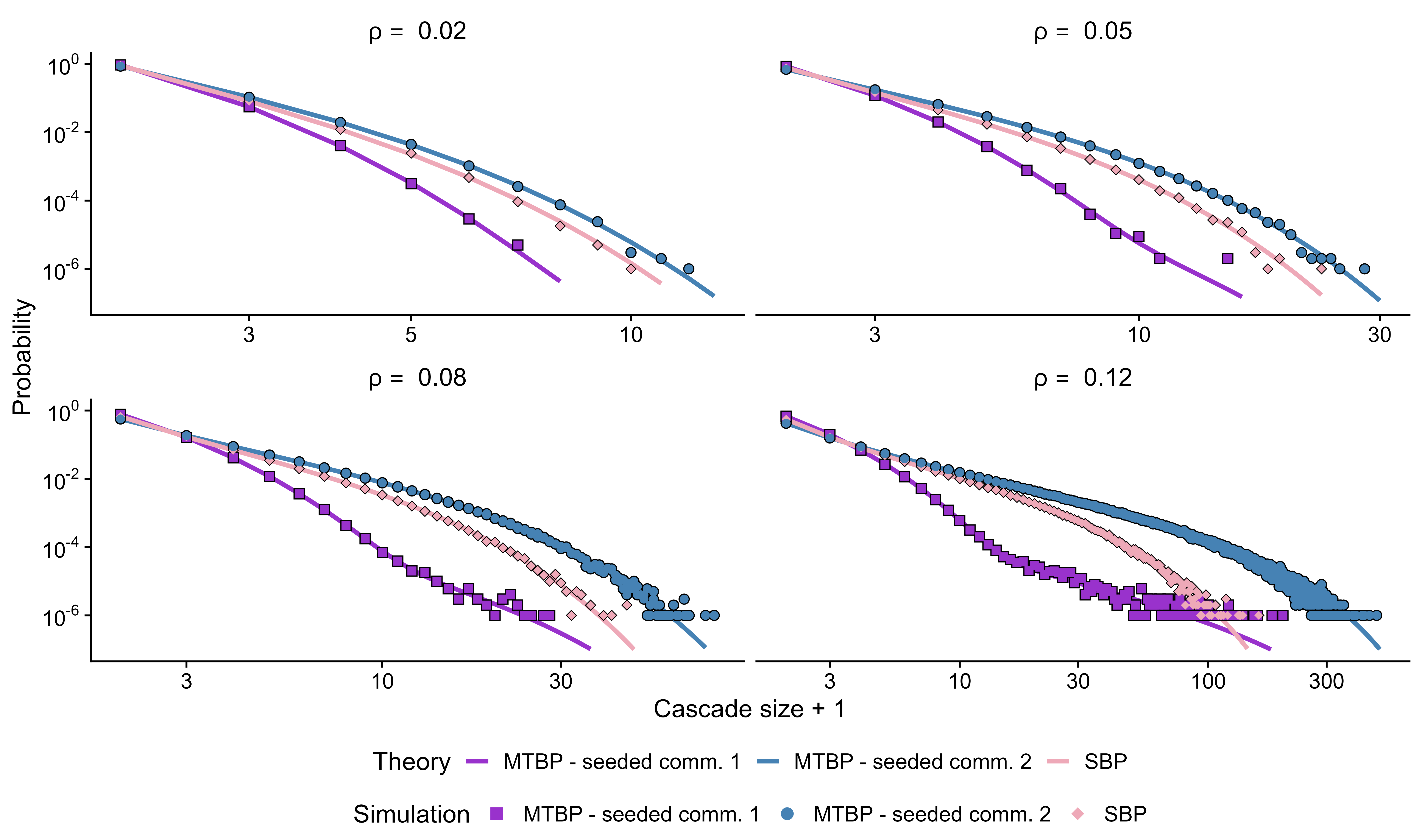}
    \caption{Cascades across the entire network when seeded in different communities for a range of infection ($\rho$) parameters. Here the cross-community connection probability, $p_e$ is fixed at $0.02$. Each community has $5\,000$ nodes, totaling $10\,000$ nodes in the network. The internal connections in community 1 and community 2 are those stated in Fig.~\ref{fig:ln_degree}. Lines represent theory curves, and the points represent the results of $N_{sim} = 10^6$ Monte Carlo simulations of the ICM model on the network. Note that the middle pink line is for the corresponding simple branching process (which does not take into account community structure).}
    \label{fig:pln_varying_rho}
\end{figure*}

This section focuses on how the community structure, in which an infection is seeded, affects the observed cascade size distribution.
Figure~\ref{fig:pln_varying_rho} shows how the cascade size distribution changes for a range of infection ($\rho$) parameters when seeded in different communities, leaving the cross-community connection probability, $p_e$, fixed at $0.02$.
We see that if we seed the branching process in community 1, which has a lower average degree, the probability of producing larger cascades is lower than if we seed initially in community 2 (which has a larger average degree). We can also note that our framework accurately captures this effect.
See App.~\ref{sec:ln_par_sweep} for a wider parameter sweep and for how these values change when we also vary the between-community connection probability $p_e$.

Next, we would like to isolate the effect of the community structure from the degree structure. To achieve this we compare the cascade dynamics on the network with communities to the cascades on the network with the same degree distribution but without community structure. To construct such a network, we take the network with community structure and perform a degree-preserving randomization. This removes the community structure but preserves the degree structure as required. We then simulate the ICM on this network to compare it to our network with community structure.

Moreover, we can easily construct a branching processes tracking diffusion on such a network and compare its prediction to the prediction of the multi-type branching process developed in this paper. To do this, we need to construct a PGF for $\mathcal{N}_s(t)$, the number of active nodes at time $t$ in a network without communities. The degree of a randomly chosen node from the network, denoted with the random variable $D_T$,  is given by
\begin{equation}
    D_T = \frac{1}{2}\overbrace{\left( D^{(1)}_1 + D^{(1)}_2\right)}^{\mathclap{\text{The total degree of nodes in community 1.}}}
    +\frac{1}{2}\underbrace{\left( D^{(2)}_1 + D^{(2)}_2\right).}_{\mathclap{\text{The total degree of nodes community 2.}}}
\end{equation}
This arises as the communities have equal size. if we randomly select a node from the entire network then we have a 50\%  chance of selecting a node from community 1 or 2. The probability generating function for the number of offspring of a randomly selected node in a network whose degree is given by the random variable $D_T$ is 
\begin{equation}
    G_{X_T}(s) = G_{D_T}\left(G_I(s)\right),
\end{equation}
where $G_{D_T}(s)=\dfrac{1}{2}\left(G_{D^{(1)}_1}(s)G_{D^{(1)}_2}(s)\right)+
\dfrac{1}{2}\left(G_{D^{(2)}_1}(s)G_{D^{(2)}_2}(s)\right)$ and 
$G_I(s) = 1-\rho + \rho s$. 
Once we are beyond the seed node, we need to consider the excess offspring distribution, $G_{\widetilde{X}_T}(s) = G^{\prime}_{X_T}(s)/G^{\prime}_{X_T}(0)$ as we did in Sec.~\ref{sec:Custom_network}. 
This allows us to write down the relationship for the number of active nodes in the network without communities, $\mathcal{N}_s(t)$, from one generation to another as  
\begin{equation}
\begin{aligned}
    G_{\mathcal{N}_s(t)}(s) &= G_{\mathcal{N}_s(t-1)}(G_{\widetilde{X}_T}(s)) \text{ for $t > 1$, and } \\   
    G_{\mathcal{N}_s(1)}(s) &=  G_{\mathcal{N}_s(0)}(G_{X_T}(s)) \text{ for $t = 1$,} 
\end{aligned}    
\end{equation}
where $G_{\mathcal{N}_s(0)}(s) = s$. The PGF   $G_{\mathcal{N}_s(1)}(s)$ provides us with a way of constructing a \textit{Simple Branching Process (SBP) } that has the same overall degree distribution as our multi-type branching process but without community structure. From this model, we can derive cascade size distributions and plot them in Fig.~\ref{fig:pln_varying_rho} alongside the cascades on the corresponding network with communities. 
From Fig. \ref{fig:pln_varying_rho}, we see that the simple branching process model for a network without communities predicts a very different cascade size distribution from the model with communities. Thus,diffusion on community-based networks cannot be captured accurately with the Simple Branching Process. The use of our MTBP framework is necessary.
\begin{figure*}[t]
    \centering
    \includegraphics[width=0.8\linewidth]{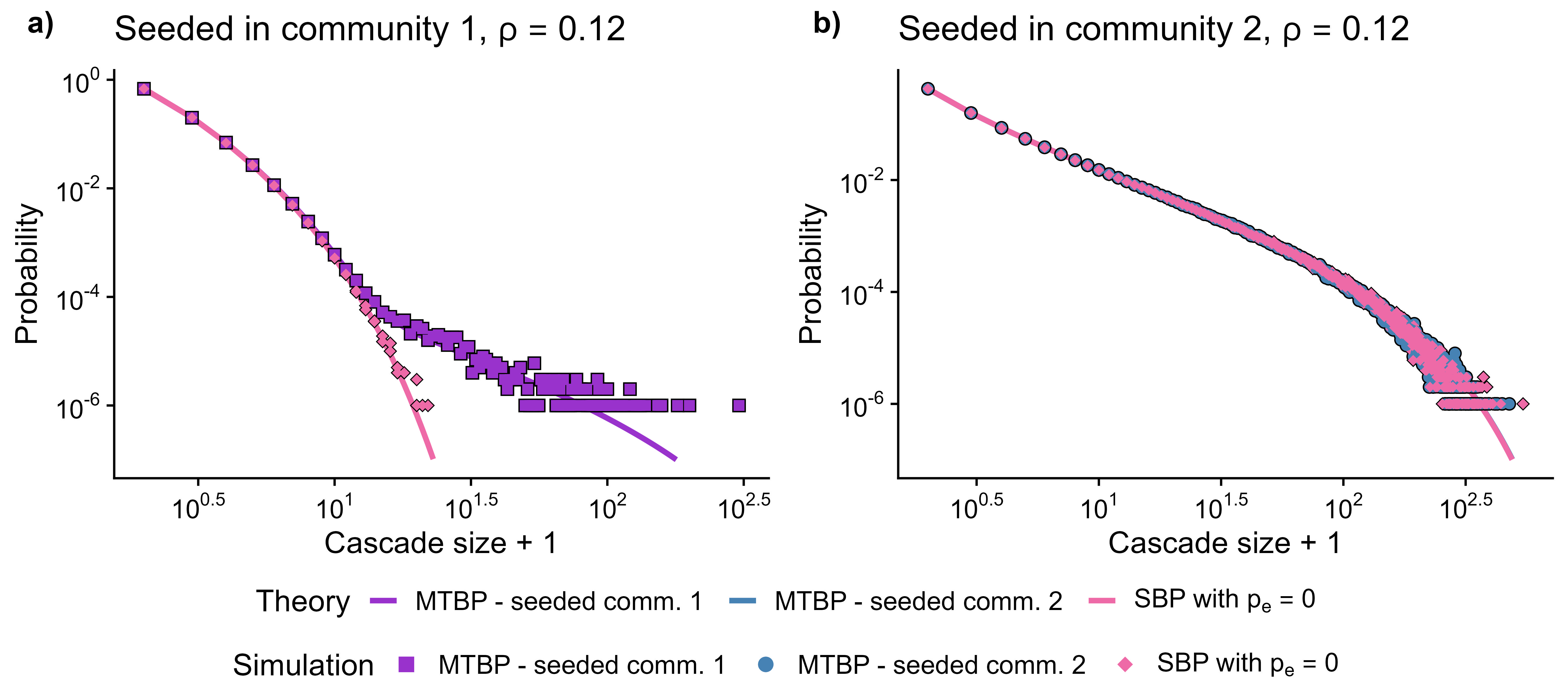}
    \caption{Cascades across the entire network when $\rho$ is fixed to $0.12$. Purple and blue lines (MTBP) represent $p_e = 0.02$, where cross-community spread is possible, and pink lines (SBP) represent $p_e = 0$, where spread is confined to one community. a) Diffusion behavior when seeding in community 1 (which has an average degree of $\approx 3$). b) Diffusion behavior when seeding in community 2 (which has an average degree of $\approx 7$). Here the blue and pink lines coincide. As before, each community has $5\,000$ nodes, totaling $10\,000$ nodes in the network. The internal connections in community 1 and community 2 use those stated in Fig.~\ref{fig:ln_degree}. Points represent the results of $N_{sim} = 10^6$ Monte Carlo simulations of the ICM model on the network. 
    }
    \label{fig:ln_effect_of_comm}
\end{figure*}

One notable effect takes place when we seed a cascade in community 1 and we have a large enough probability of infection, $\rho = 0.12$, the chance of observing larger cascades changes and displays quantitatively similar behavior to those cascades seeded in community 2 (i.e., the shape deviates and appears similar to cascade size distribution when cascades are seeded in community 2). We can speculate that this is due to activation probability being large enough that, if we observe a cascade of a certain size, it is likely to have spread to community 2, where the process takes advantage of the larger average degree within that community. In contrast, we see no similar effect in community 2. 

We can easily isolate the effect of communities by removing the inter-community links, setting $p_e=0$ to verify our interpretations. This allows the activations to spread within either community 1 or 2 only. Fig.~\ref{fig:ln_effect_of_comm}a) shows the cascade size distribution when information is seeded within community 1 and either allowed to spread between communities or not, while Fig.~\ref{fig:ln_effect_of_comm}b) shows the cascade size distribution when information is seed within community 2 and either allowed to spread between communities or not.  We can note from Fig.~\ref{fig:ln_effect_of_comm}a) that it is indeed due to the community structure that community 1 sees a change in its cascade size distribution, where the inter community links provide a means of linking to the higher, on average, degree nodes in community 2. However, if we seed activations in community 2, we see no such discernible benefit for community 2 being a connection to community 1, where the inter community links provide a means of linking to the lower, on average, degree nodes in community 1, which does not have a qualitative effect on the cascade size distributions shape. The interested reader can see the effect of communities whose average degree is closer to each others in App. \ref{app:MTBP_degree_4_5}.

In this section, we have shown that our multi-type branching process model for diffusion between communities can accurately model the spread of entities between communities with a general network structure. This captures not only the effect of the community structure, where we have different internal connection structures within the communities, but also the effect of seeding infection in one community over the other. In the following section we provide some conclusions of our work and future directions. 

\section{Conclusion}\label{sec:Conclusion}

In this paper, we studied the spreading process on a broad class of networks with community structures. We showed how, by using limited information about the network, specifically the degree distribution within and between the communities, we can calculate standard statistical characteristics of cascade dynamics. For this purpose, we developed a probability generating function theory to study how the diffusion process spreads between and within communities. 
The methods developed in this study add to the theoretical tools for studying diffusion on community-based networks or multiple connected networks.

Our framework allowed us to calculate important quantities of the diffusion process, such as extinction probability, hazard function, and cascade size distribution, proving to be a simple and effective tool. Not only did this tool enable us to accurately estimate the properties of dynamics on the entire network, but it also allowed us to extend these quantities to community-specific versions of characteristics of interest. This allowed us, for instance, to isolate the probability of extinction in a specific community. Additionally, we derived new quantities, specific to diffusion on a network with communities, such as the introduction and reintroduction probabilities. The introduction probability estimates (with high accuracy) the probability that the infection spreads to a new community, while the reintroduction represents the probability that a contagion, having ceased spreading in a specific community, might be reintroduced to that community at a later stage.

Finally, we analyzed how the diffusion process spreads across and within communities in a particular case in more details. We focused on a network that can be used to represent social networks, which was a graph with heavy-tailed (log-normal) distributed degree and sparsely distributed inter-community edges. We demonstrated how the initial seeding location of an infection affects the observed cascade size distribution.
Furthermore, we compared the cascade dynamics on a network with communities to the dynamics on a network with the same degree distribution but no communities present. We showed that the presence of communities notably change the cascade dynamics on a heavy-tailed network and that our framework accurately captures the changes.

In this work, we have used the Independent Cascade Model (ICM) as the diffusion process. Although very useful for biological contagions, it may not describe how information spreads on a real social network. In \cite{gleeson_branching_2021}, it was noted that the empirical cascade size distribution of social spreads was not captured well by the ICM. However, simple extensions of the ICM performed very well. For instance, the Limited Attention Model \cite{lerman2016information, weng2012competition} scales the probability of a node spreading a piece of information by the number of links they have. This scaling attempts to capture the cognitive load of users who may see many pieces of content from many users that they are in contact with. Such a model would be a natural extension to this work, which would better reflect real-world spread dynamics than the ICM and is an interesting future direction.
 
While this paper only discusses networks with two communities, our framework can be easily extended to accommodate any number of communities. This does not pose any theoretical challenges; the only complication will be the rapid increase in the number of random variables that need to be tracked as the number of communities increases. There are also several other directions that could be explored in future work. One natural extension would be incorporating a clique structure into the network within our framework. In the present paper, we require the network to have a local tree-like structure. The methods developed in~\cite{keating_multitype_2022, Keating_Gleeson_OSullivan_2023} will be a good starting point for this. Additionally, in the continuation of developing theoretical tools, it seems feasible to extend our methods to directed networks and networks with degree-degree correlations.

To further investigate the effect of communities on propagation dynamics, it may be interesting to study cases involving communities of varying sizes and determine whether the size of a community influences cascade dynamics, as it reasonably would. Another intriguing scenario involves introducing different types of nodes with varying probabilities of infection. This addition will not significantly complicate our analysis but will increase the number of random variables we need to track within our framework.

To enhance the applicability of our method described here, a comparison to real world data is a natural extension. Previous work~\cite{pena2023finding} tracked information diffusion between and within communities for two conversation networks on Twitter (now X). The authors showed that users tend to communicate more frequently to other users in the same community than to users in the opposite community. But to what extent this is explained by the network structure and diffusion processes is an interesting and open question. The methodology introduced in this paper opens the door to the application to empirically observed cascades, to determine to what extent the network structure can explain the observed information diffusion process.

\section*{Funding}\label{sec:Acknowledgement}

This publication has emanated from research conducted with the financial support of Science Foundation Ireland under Grant numbers 18/CRT/6049 and 16/IA/4470, and European Research Council under the European Union’s Horizon 2020 research and innovation programme grant number 802421. For the purpose of Open Access, the authors have applied a CC BY public copyright licence to any Author Accepted Manuscript version arising from this submission.
\newpage

\appendix
\section{Numerical estimation of probabilities from probability generating functions}\label{sec:FFT}

Let us first consider the case of the univariate probability-generating function
\begin{equation}
G_{N(t)}(s) = \sum_0^\infty P[N(t)=n] s^n. 
\end{equation}
By definition, the probability $P[N(t) = n]$ is recovered from the PGF $G_{N(t)}(s)$ as

    \begin{equation}
        P[N(t)] = \frac{1}{n!}\left[ \frac{d^{n}}{d s^{n}}\left( G_{N(t)}(s) \right) \right]_{(s=0)}.
    \end{equation}
By Cauchy's theorem, we can transform this into a contour integral in the complex plane~\cite{Cavers_1978, newman2001random} as 

\begin{equation}
    P[N(t) = n] = \frac{1}{2\pi i} \oint_C \frac{G_{N(t)}(s)}{s^{(n+1)}} d s,
\end{equation} 
where we can choose the unit circle for the contour $C$~\cite{newman2001random}. Next, replacing $s=e^{i\omega}$, we obtain 

\begin{equation}
    P[N(t) = n] = \frac{1}{2\pi}\int_0^{2\pi} G_{N(t)}(e^{i\omega}) e^{-i n\omega} d\omega.
    \label{eq:app0_1}
\end{equation} 
To calculate Eqn. \eqref{eq:app0_1} numerically, we use the trapezoidal rule and divide the unit circle into $K$ points. This yields

\begin{equation}
    P[N(t) = n] \approx \frac{1}{K} \sum_{k=0}^{K-1} G_{N(t)}(e^{2\pi i k/K}) e^{2\pi i n k/K}.
    \label{eq:app0_2}
\end{equation} 
In turn, equation \eqref{eq:app0_2} can be evaluated using standard fast Fourier transform numerical techniques~\cite{Cavers_1978, marder2007dynamics}.

In the case of the bivariate PGF, the joint probability $P[N_1(t) = n, N_2(t) = m]$ can be recovered from the bivariate PGF, $\G{N}{t}(s_1, s_2)$, as
\begin{widetext}
    \begin{equation}
        P[N_1(t) = n, N_2(t) = m] = \frac{1}{n! m!}\left[ \frac{\partial^{n}}{\partial s_1^{n}}\frac{\partial^{m}}{\partial s_2^{m}} \left( \G{N}{t}(s_1, s_2) \right) \right]_{(s_1=0,s_2=0)}.
        \label{eq:app0_2_1}
    \end{equation}
\end{widetext}
Similarly to the univariate case, if we substitute $s_1=e^{i\omega}$ and $s_2=e^{i\theta}$ into the expression for $\G{N}{t}(s_1, s_2)$, we can use 2-dimensional discrete Fourier transform to evaluate probabilities for the first $K$ values for $N_1(t)$ and $N_2(t)$. This yields

\begin{widetext}
\begin{equation}
    P[N_1(t) = n, N_2(t) = m] \approx \frac{1}{K^2} \sum_{k_1=0}^{K-1} \sum_{k_2=0}^{K-1} \G{N}{t}(e^{2\pi i k_1/K}, e^{2\pi i k_2/K}) e^{2\pi i n k_1/K}e^{2\pi i m k_2/K}.
    \label{eq:app0_3}
\end{equation} 
\end{widetext}

\section{Recovering probabilities from probability generating functions}\label{sec:appendix_recovering_probabilities}

In this section we give detailed explanation on how the probabilities are recovered from the corresponding probability generating functions. As explained in Sec.~\ref{sec:pgf_framework}, the probability generating function, $\G{N}{t}$, by definition is
\begin{equation}
    \G{N}{t}(s_1,s_2) = \sum_{n=0}^\infty \sum_{m=0}^\infty P[N_1(t)=n,N_2(t)=m] s_1^n s_2^m.
    \label{eq:PGF_def_appendix}
\end{equation}
As explained in Appendix~\ref{sec:FFT}, the joint probability $P[N_1(t) = n, N_2(t) = m]$ is given by the derivative of $\G{N}{t}$, as defined by \eqref{eq:app0_2_1}, and is evaluated using the discrete Fourier transform given by Eqn.~\eqref{eq:app0_3}.

Now, let us discuss how we recover the probability of the number of active nodes in community 1 only. If we substitute $s_2=1$ into Eqn.\eqref{eq:PGF_def_appendix} we obtain 

\begin{equation}
    \G{N}{t}(s_1,1) = \sum_{n=0}^\infty P[N_1(t)=n] s_1^n.
\end{equation}
and the probability $P[N_1(t)=n]$ is straightforwardly recovered by
\begin{equation}
    P[N_1(t)=n] = \frac{1}{n!}\left[\frac{d^n}{d s_1^n}\left(\G{N}{t}(s_1,1)\right) \right]_{s_1=0}.
\end{equation}
To calculate this probability numerically, we apply discrete Fourier transform to function $\G{N}{t}(s_1,1)$ as follows
\begin{equation}
    P[N_1(t)=n] \approx \frac{1}{K} \sum_{k=0}^{K-1} G_{N(t)}(e^{2\pi i k/K},1) e^{2\pi i n k/K}.
    \label{eq:p_n_appendix}
\end{equation} 
Similarly, we apply discrete Fourier transform to $\G{N}{t}(1,s_2)$ to evaluate the probability of the number of active nodes in community 2, $P[N_2(t)=n]$. 

Finally, let us discuss how to obtain the probability of the total number of active nodes in the entire network, $P[N_1(t)+N_2(t)=n]$. To do this, we substitute $s_1=s$ and $s_2=s$ into $\G{N}{t}$ to obtain

\begin{equation}
    \G{N}{t}(s,s) = \sum_{n=0}^\infty \sum_{m=0}^\infty P[N_1(t)=n,N_2(t)=m] s^{n+m}.
\end{equation}
Then the probability, $P[N_1(t)+N_2(t)=n]$, is calculated as follows
\begin{equation}
    P[N_1(t)+N_2(t)=k] = \frac{1}{k!}\left[\frac{d^k}{d s^k}\left(\G{N}{t}(s,s)\right) \right]_{s=0}.
\end{equation}
which, in its turn, is evaluated with the discrete Fourier transform of $\G{N}{t}(s,s)$ as follows
\begin{align}
    P[N_1(t)+& N_2(t)=n] \\ \approx &\frac{1}{K} \sum_{k=0}^{K-1} G_{N(t)}(e^{2\pi i k/K},e^{2\pi i k/K}) e^{2\pi i n k/K}.
\end{align} 

\newpage
\section{Comparing branching process simulations to network simulations}\label{sec:Simulations_BP_Network}

We compare our model using multi-type branching processes to network-based simulations and to the expected diffusion using the mean-matrix to check the accuracy of the model. The mean-matrix $M$ tracks the expected offspring of each type, so $M_{ij}$ is the expected number of offspring from a node $i$ that are of type $j$ in the next time step. For the SBM we have the mean-matrix

$$
M = 
\begin{pmatrix}
    \rho \lambda_{in} & \rho \lambda_{out} \\ \rho \lambda_{out} & \rho \lambda_{in}
\end{pmatrix}.
$$

If we start from a single seed in community 1, we can track the expected number of nodes by generation $t$  in community 1 as 
$$
E[N_1 (t)]= \begin{pmatrix}
    1 & 0
\end{pmatrix} M ^ t 
\begin{pmatrix}
    1 \\ 0
\end{pmatrix},
$$
where $t$ denote the matrix power and generation $t$. Similarly, if we are interested in tracking the number of active nodes in community 2 by generation $t$ from a single seed in community 1 then we have 
$$
E[N_2 (t)]= \begin{pmatrix}
    1 & 0
\end{pmatrix} M ^ t 
\begin{pmatrix}
    0 \\ 1
\end{pmatrix}.
$$

We can then use this to compare our network-based simulations to the theory with the mean-matrix and to the branching processes simulations. We will see that all three have strong agreement in the subcritical case and for sufficiently large networks. Therefore, in some places, we will use branching processes simulations thanks to their ease of simulation.

\begin{figure*}
    \centering
    \includegraphics[width = 0.8\textwidth]{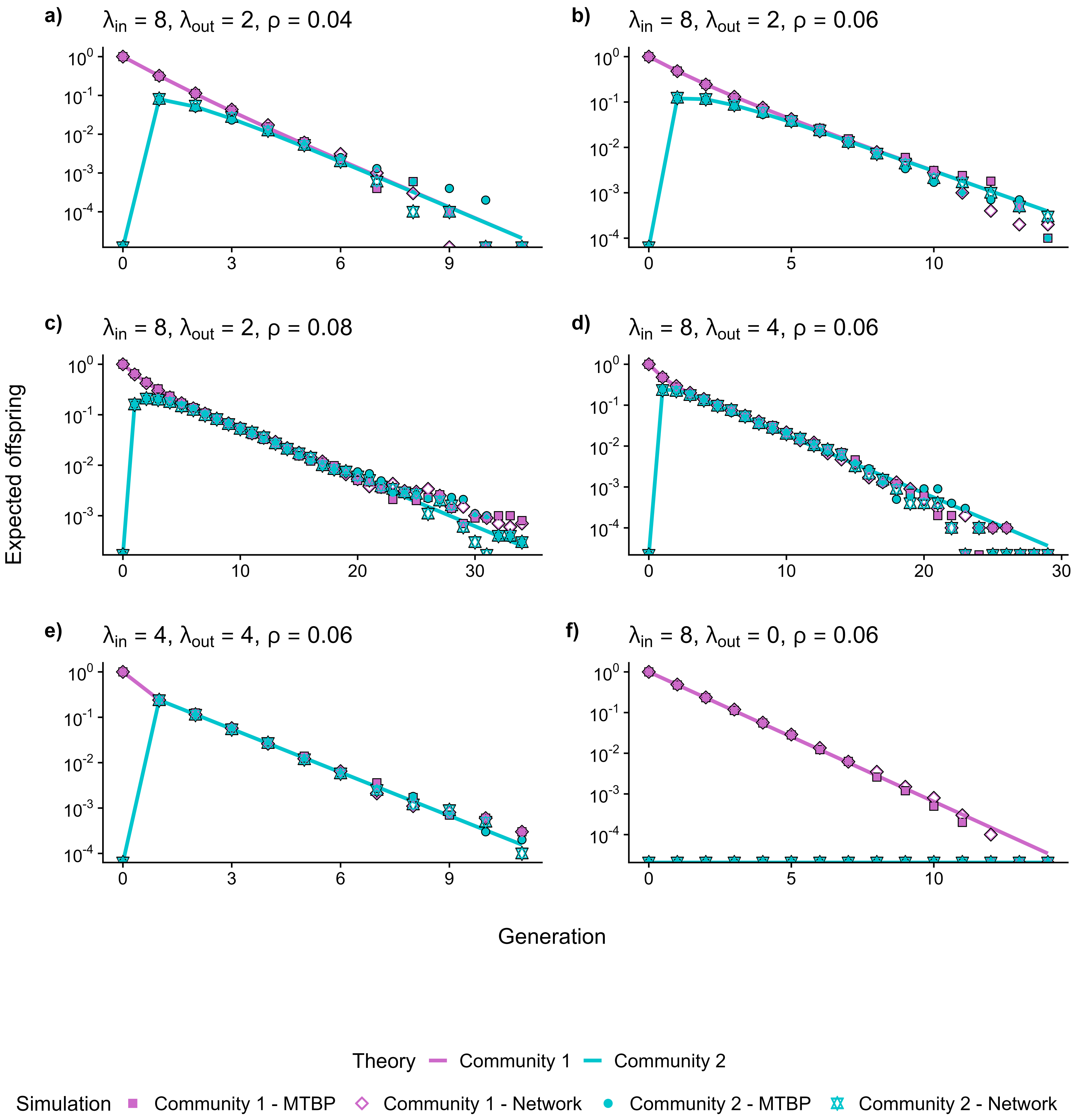}   
    \caption{Distribution of expected offspring for various parameters. The shapes account for the multi-type branching process (MTBP) simulations and the network simulations, and the lines refer to the results given by the mean-matrix. a) $\lambda_{\text{in}} = 8$, $\lambda_{\text{out}} = 2$, $\rho = 0.04$: information spreads shortly overall; b) $\lambda_{\text{in}} = 8$, $\lambda_{\text{out}} = 2$, $\rho = 0.06$: moderate information spread overall; c) $\lambda_{\text{in}} = 8$, $\lambda_{\text{out}} = 2$, $\rho = 0.08$: long diffusion process overall;  d) $\lambda_{\text{in}} = 8$, $\lambda_{\text{out}} = 4$, $\rho = 0.06$: information spreads further than in (b) but dies out slightly earlier than in (c), with moderate diffusion overall; e) $\lambda_{\text{in}} = \lambda_{\text{out}} = 4$, $\rho = 0.06$: information dies out early, comparable to (a); f) $\lambda_{\text{in}} = 8$, $\lambda_{\text{out}} = 0$, $\rho = 0.06$: information spreads only in community 1, with moderate diffusion.}
    \label{fig:total_size}
\end{figure*}

Figure~\ref{fig:total_size} shows the results obtained for $10\,000$ simulations for a range of values of infection probability $\rho$ ($\rho = 0.04$, $\rho = 0.06$, $\rho = 0.08$), expected number of edges inside a community $\lambda_{\text{in}}$ ($\lambda_{\text{in}} = 4$, $\lambda_{\text{in}} = 8$) and expected number of edges between communities $\lambda_{\text{out}}$ ($\lambda_{\text{out}} = 0$, $\lambda_{\text{out}} = 2$, $\lambda_{\text{out}} = 4$) on a Poisson-distributed network. We use Stochastic Block Model networks containing $10\,000$ nodes split evenly between community 1 community 2 and compare those to the equivalent multi-type branching process model for Poisson distribution. The simulations were set to start with the activation of one random node in community 1. There is cross-community spread when $\lambda_{\text{out}} > 0$, in which case community 2 is expected to be infected on the first passage, on average. Fig.~\ref{fig:total_size}c) shows the offspring distribution for $\rho = 0.08$, $\lambda_{\text{in}} = 8$ and $\lambda_{\text{out}} = 2$. This is the combination of parameters among the considered values that generates the longest process. The process dies off earlier for the simulations using $\rho = 0.06$ (Fig.~\ref{fig:total_size}b)) and $\rho = 0.04$ (Fig.~\ref{fig:total_size}a)) since the infection survives shortly when the probability of infection is lower. When we consider no possibility of cross-community diffusion ($\lambda_{\text{out}} = 0$, Fig.~\ref{fig:total_size}f)), we observe diffusion only inside community 1, and the overall process dies off earlier than the analogous with $\lambda_{\text{out}} = 2$ (Fig.~\ref{fig:total_size}b)). On the other hand, when we increase $\lambda_{\text{out}}$ (Fig.~\ref{fig:total_size}d)) infection spreads further in both community 1 and community 2 in comparison to the analogous with lower $\lambda_{\text{out}}$ values (Fig.~\ref{fig:total_size}b) and f)). When we keep $\lambda_{\text{out}}$ moderate ($\lambda_{\text{out}} = 4$) and decrease $\lambda_{\text{in}}$ ($\lambda_{\text{in}} = \lambda_{\text{out}} = 4$), with moderate probability of infection, the process dies out early.  

Both the multi-type branching process and network-based simulations show good agreement with the theoretical expected results, suggesting that our model is accurate for large networks.

\newpage
\section{Log-normal network parameter sweep}\label{sec:ln_par_sweep}

In this section we provide a parameter sweep for both $\rho$ and $p_e$, and analyze the resultant distributions of cascade sizes. The network is the same as that presented in Sec.~\ref{sec:Custom_network}, namely where we have two communities of $5\,000$ nodes, each with a log-normal distribution internally. Community 1 has $\mu_1 = 0.85$ and $\sigma_1 =0.4$, resulting in an internal  average degree of 3. Community 2 has $\mu_2 = 1.8$ and $\sigma_2 = 0.4$, resulting in an internal average degree of 7. The cross-community links are selected by adding single edges between nodes. We randomly select one node from community 1 and one node from community 2 and link them. The proportion of nodes selected is $p_e$. 

\begin{figure*}[t]
    \centering
    \includegraphics[width = 0.85\textwidth]{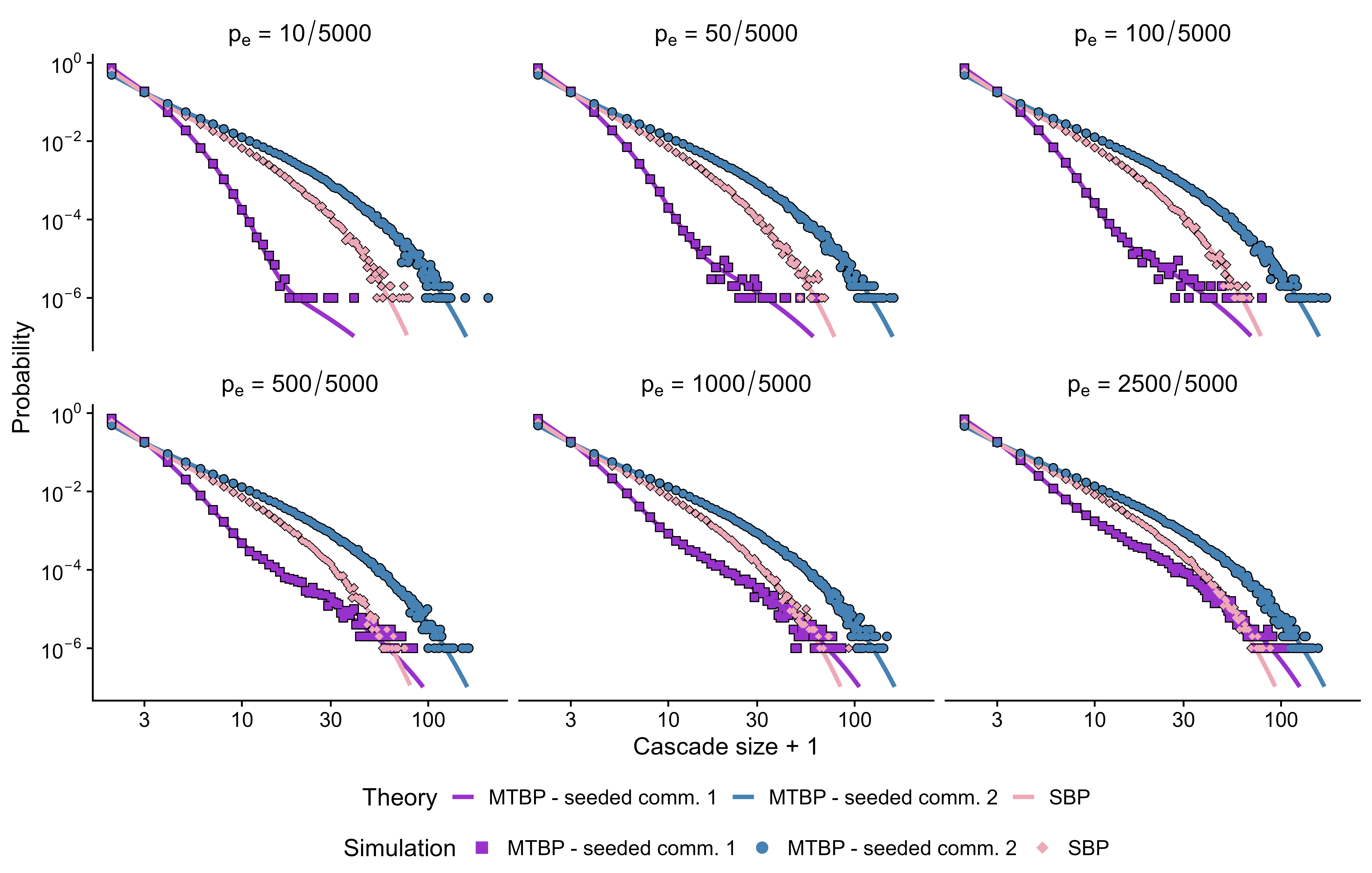}   
    \caption{Parameter sweep, where $\rho$ is fixed at $0.1$ and $p_e$ is swept from $10/5\,000$ to $2\,500/5\,000$. Each community has $5\,000$ nodes, totaling $10\,000$ nodes in the network. Lines represent theory curves, and the points represent the results of $N_{sim} = 10^6$ Monte Carlo simulations of the ICM model on the network. Note that the middle pink line is for the corresponding simple branching process (which does not take into account community structure).}
    \label{fig:app_pe_sweep}
\end{figure*}

Figure~\ref{fig:app_pe_sweep} shows the comparison of the theory and simulation under different values of $p_e$, where $\rho$ is fixed. We can see that we have excellent agreement between our theory and simulations. As we increase the number of cross-community edges we have a general increase in cascade size. This is most pronounced for those cascades seeded in community 1 (purple line).

\begin{figure*}
    \centering
    \includegraphics[width = 0.85\textwidth]{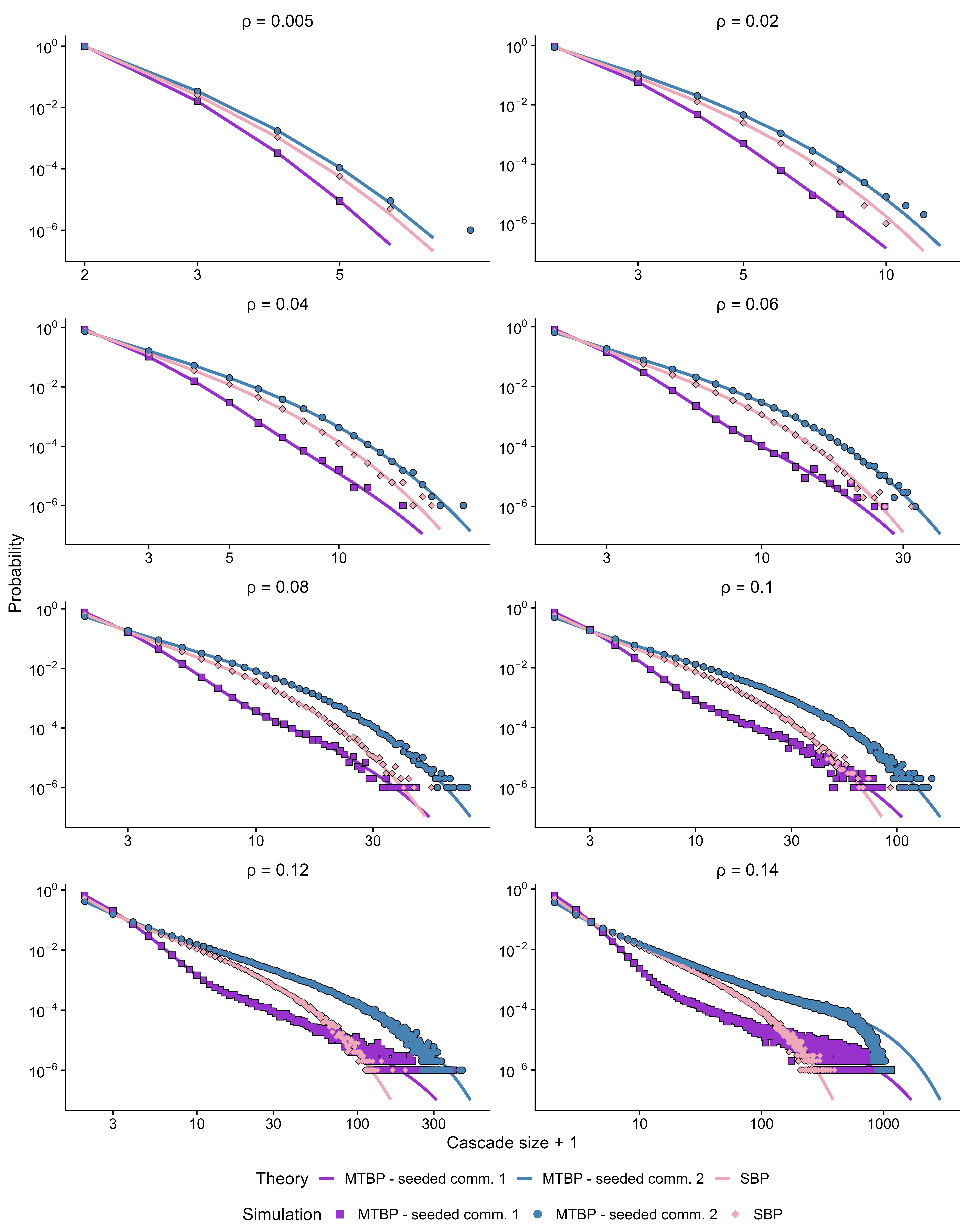}   
    \caption{Parameter sweep, where $p_e$ is fixed at $1\,000/5\,000$ and $\rho$ is swept from $0.005$ to $0.14$. Each community has $5\,000$ nodes, totaling $10\,000$ nodes in the network. Lines represent theory curves, and the points represent the results of $N_{sim} = 10^6$ Monte Carlo simulations of the ICM model on the network. Note that the middle pink line is for the corresponding simple branching process (which does not take into account community structure).}
    \label{fig:app_pinf_sweep}
\end{figure*}

Figure~\ref{fig:app_pinf_sweep} provides the behavior of our cross-community branching processes for various $\rho$ values, where $p_e$ is fixed to $1\,000/5\,000$. We can note that we have excellent agreement between our theory and simulations in general. For larger $\rho$ (most pronounced for $\rho = 0.14$) values, we can note discrepancies between the theory and simulated values. The larger simulated cascade do not reach the same sizes as the theoretical cascades sizes. This is due to the finite size of the network, as for this value, cascades activate a significant proportion of the networks, while in the theory it is assumed that the networks are infinitely large.

As shown in Fig.~\ref{fig:app_pinf_sweep}, cascades seeded in the second community appear to follow a power-law distribution. Table~\ref{tab:alpha_ext} below illustrates the power-law exponent ($\alpha$) derived using a maximum likelihood approach with the simulated cascade data for multiple infection probabilities ($\rho$). Notably, as the infection probability increases and the process nears the critical point, the characteristic exponent approaches 
$3/2$, which is consistent with expectations for the critical regime.

Estimating the power-law exponent for cascades seeded in the first community is more complex due to a sharp transition between two regimes in their cascade size distribution. The cascades modeled using the SBP in the figure primarily serve to demonstrate that the SBP model is not well-suited for diffusion processes on networks with communities. As such, these results were not pursued further.
It is important to note that these results are derived from the simulation data. Ideally, we would like to derive asymptotic expression for the cascade size distribution to predict this scaling; this is left to further work.


\begin{table}[h]
\begin{tabular*}{\columnwidth}{@{\extracolsep{\fill}}ccccc}
\hline
\multicolumn{5}{c}{Cascades seeded in community 2} \\ 
\hline
$\rho$  & 0.08 & 0.10 & 0.12 & 0.14 \\ 
$\alpha$ & 1.83 & 1.72 & 1.60 & 1.54 \\ 
\hline
\end{tabular*}
\caption{Estimates of the scaling parameter $\alpha$ for different infection probabilities $\rho$ obtained using the maximum likelihood estimator method~\cite{newman_networks_2018}. All other simulation parameters are the same as those in Fig.~\ref{fig:app_pinf_sweep}.}
\label{tab:alpha_ext}
\end{table}
\newpage
\section{Constructing a multi-type branching process for k-regular networks}\label{sec:MTBP_simple_example}

In Sec.~\ref{sec:Custom_network}, we derived the required equations that describe the spread dynamics under the ICM through a network with two communities. Later, in Sec.~\ref{sec:ln_network_example}, we provided an example of how this works for a log-normally distributed network. As an aid to the reader, we provide here a simplified network example. We will assume the internal structure of the communities will be $k$-regular. Each node in community 1 will have in-community degree of $3$ and each node in community 2 will have in-community degree of $7$. As before, the probability of a node becoming active from an exposure in the previous timestep is $\rho$. The allows us to write down our PGF for the internal degree of each node in both communities, and their offspring distributions 
\begin{align*}
    G_{\dis{D_1}{1}}(s) 
    & = s^3 \\
    \Rightarrow G_{\dis{X_1}{1}}(s) 
    &= G_{\dis{D_1}{1}}(G_I(s)) \\
    &= (1-\rho + \rho s)^3 \text{ and } \\
    G_{\dis{D_2}{2}}(s) &= s^7 \\
     \Rightarrow G_{\dis{X_2}{2}}(s) 
     &= G_{\dis{D_2}{2}}(G_I(s))\\
     &= (1-\rho + \rho s)^7, \\
\end{align*}
from which we can derive the required excess distributions using Eq.~\eqref{eq:pgftoexcesspfg} as 
\begin{align*}
    G_{\dis{\widetilde{X}_1}{1}}(s) 
    &= G_{\dis{\widetilde{D}_1}{1}}(G_I(s))
    = (1-\rho + \rho s)^2 \text{ and } \\
    G_{\dis{\widetilde{X}_2}{2}}(s) 
    &= G_{\dis{\widetilde{D}_2}{2}}(G_I(s))
    = (1-\rho + \rho s)^6. \\
\end{align*}
We assume, as we did in Sec.~\ref{sec:ln_network_example}, that each node has probability $p_e$ of having $1$ link between the communities, 
\begin{align*}
G_{\dis{X_{2}}{1}}(s) = G_{\dis{X_{1}}{2}}(s)  \\
            &= 1-p_e + p_e (G_I(s)) \\
            &= 1-p_e + p_e (1 - \rho + \rho s), \nonumber    
\end{align*}
from which we can easily show that the excess offspring distributions for between-community spread are given by $G_{\dis{\widetilde{X}_{2}}{1}}(s) = G_{\dis{\widetilde{X}_{1}}{2}}(s) = 1$. Now we can proceed to create the offspring distribution for each of the 4 types that we have to track in our branching process. First, let us consider the PGF for the offspring from a node in community 1, who was activated from within its community. As it was activated from inside community 1, it will have one less neighbor that it can activate and this requires the use of the excess distribution, which is associated random variable $\dis{\widetilde{X}_1}{1}$, while no cross-community links we used in the activation, and so will use the cross-community degree distribution, which is associated with the random variable $\dis{X_2}{1}$, yielding the following offspring distribution:
\begin{align*}
    \g{X_{in}}{1}(&s_1^{in},s_2^{out})  
    = G_{\dis{\widetilde{X}_1}{1}}(s_1^{in})G_{\dis{X_2}{1}}(s_2^{out}) \\
    &= (1-\rho + \rho s_1^{in})^2 \left( 1-p_e + p_e (1 - \rho + \rho s_2^{out}) \right) .\\
\end{align*}
When deriving $\g{X_{out}}{1}$ the same logic holds, but as the node was activated by from outside of community 1, therefore, it will still have its full complement of in-community nodes that it can activate, requiring the use of $\dis{X_1}{1}$ and the use of the excess between-community offspring distribution, with associated random variable $\dis{\widetilde{X}_2}{1}$. The yielding the following offspring distribution 
\begin{align*}
    \g{X_{out}}{1}(s_1^{in},s_2^{out}) 
    &= G_{\dis{X_1}{1}}(s_1^{in})G_{\dis{\widetilde{X}_2}{1}}(s_2^{out}) \\
    &= (1-\rho + \rho s_1^{in})^3 \left( 1 \right) . \\
\end{align*}
And we can easily extend this to find the offspring distributions for community 2, which will be  
\begin{align*}
    \g{X_{in}}{2}(&s_2^{in},s_1^{out}) 
    = G_{\dis{\widetilde{X}_2}{2}}(s_2^{in})G_{\dis{X_1}{2}}(s_1^{out}) \\
    &= (1-\rho + \rho s_2^{in})^6 \left( 1-p_e + p_e (1 - \rho + \rho s_1^{out}) \right) \text{ and } \\
    \g{X_{out}}{2}(&s_2^{in},s_1^{out}) 
    = G_{\dis{X_2}{2}}(s_2^{in})G_{\dis{\widetilde{X}_2}{1}}(s_2^{out}) \\
    &= (1-\rho + \rho s_2^{in})^6 \left( 1 \right) . \\
\end{align*}
To find the number of nodes active at time $t$ we just need to consider what happens between generation 0 and 1. If we seed the process with one active individual in community 1 at time $t=0$, then our initial conditions are $G_{\mathcal{N}(1)}(s) = (s)^1$. Then as the process has just started the active node has no previously active neighbors, therefore, the offspring distribution is given by 
\begin{align*}
    G_{\mathcal{N}(1)}(&s_1^{in}, s_1^{out},s_2^{in}, s_2^{out}) 
    = \g{X}{1}(s_1^{in}, s_1^{out},s_2^{in}, s_2^{out}) \\
    &= G_{\dis{X_1}{1}}(s_1^{in})G_{\dis{X_2}{1}}(s_2^{out}) \\
    &= (1-\rho + \rho s_1^{in})^3 \left( 1-p_e + p_e (1 - \rho + \rho s_2^{out})\right) .
\end{align*}

Now we can iterate this function to the desired number of generations, for example, to find the number of active nodes by generation $2$, we can note that 

\begin{widetext}
\begin{align*}
    G_{\mathcal{N}(2)}(s_1^{in}, s_1^{out},s_2^{in}, s_2^{out}) &= 
    G_{\mathcal{N}(1)}(\g{X_{in}}{1}, \g{X_{out}}{1},\g{X_{in}}{2},\g{X_{out}}{2}) \\
    &= (1-\rho + \rho (\g{X_{in}}{1}))^3 
    \left( 1-p_e + p_e (1 - \rho + \rho (\g{X_{out}}{2}))\right) \\
    &= (1 - \rho  + \rho  (1 - \rho  + s_1^{in} \rho )^2 (1 - p_e + (1 - \rho  + s_2 \rho) p_e))^3 \times \\
    &\qquad\qquad (1 - p_e + (1 - \rho + \rho  (1 - \rho  + s_2^{in} \rho )^7) p_e).
\end{align*}
\end{widetext}

This process can be continued to find later generations, which, for small $t$ can be found easily using symbolic solvers; however, for a later generations the use of symbolic solvers becomes computationally expensive requiring methods of inverting probability generations need to be used, which we have discussed in App. \ref{eq:app0_1}. We also make use of additional types when we wish to calculate a probability generating function for the total cascade size. Effectively, this new particle type is generated every time a node is actives, but they themselves never die. This addition of counter types results in the ability of to count the number of active node created by time $t$. Let $c$ be a dummy variable for this new type, in the last example our initial conditions would now be, 

$$
G_{\mathcal{C}(0)}(s, c) = (s)^1 c.
$$
Then the $t=1$ generation would be
\begin{align*}
    G_{\mathcal{C}(1)}(&s_1^{in}, s_1^{out},s_2^{in}, s_2^{out},c,c) 
    = \g{X}{1}(s_1^{in} c, s_2^{out} c,c) \\
    &= G_{\dis{X_1}{1}}(s_1^{in} c)G_{\dis{X_2}{1}}(s_2^{out} c)c \\
    &= (1-\rho + \rho s_1^{in} c)^3 \left( 1-p_e + p_e (1 - \rho + \rho s_2^{out} c)\right) c.
\end{align*}
 We iterate similarly to the previous example to arrive at the required PGF 
\begin{align*}
    &G_{\mathcal{C}(2)}(s_1^{in}, s_1^{out},s_2^{in}, s_2^{out},c,c) 
    \\&=  (1 -\rho + c\rho (1 -\rho + c s_1^{in} \rho)^2 \times 
    \\ &\qquad (1 - p_e + (1 -\rho + c s_2^{out} \rho)  p_e))^3 \times 
    \\ &\qquad\qquad (1 - p_e + (1 -\rho + c\rho (1 -\rho + c s_2^{in}\rho)^7) p_e) c,
\end{align*}
and this process is repeated until the required generation has been reached. We can then marginalized the required types in the PGF to recover the univariate PGF for the cascade size distribution, for example, by $G_{\mathcal{C}(t)}(1,1,1,1,c)$. Then this can be inverted via inverse fast Fourier transform to obtain the probability mass function, see App. \ref{eq:app0_1}. It should be noted that this entire process is no more complex for general degree distributions once we have the network structure, just a little more cumbersome. 
\newpage
\section{The community seeding effect on a log-normal network with communities of similar degree distribution}\label{app:MTBP_degree_4_5}

This section repeats the experiments from Sec. \ref{sec:effect_community}, where we isolated the effect of inter community spreading by comparing two communities with approximately log normally degree distribution, where community 1 had an average degree of approximately 3 and community 2 had an approximate degree of 7. We were able to note in Fig. \ref{fig:ln_effect_of_comm} that community 1, with the lower average degree, had a qualitatively different cascade distribution when we allowed inter community spread; whereas community 2 had no such discernible effect. Here we repeat this comparison but now using two communities with closer average degrees; now community 1 has an average degree of 4 and community 2 has an average degree of 5.

\begin{figure*}[t]
    \centering
    \includegraphics[width=0.8\linewidth]{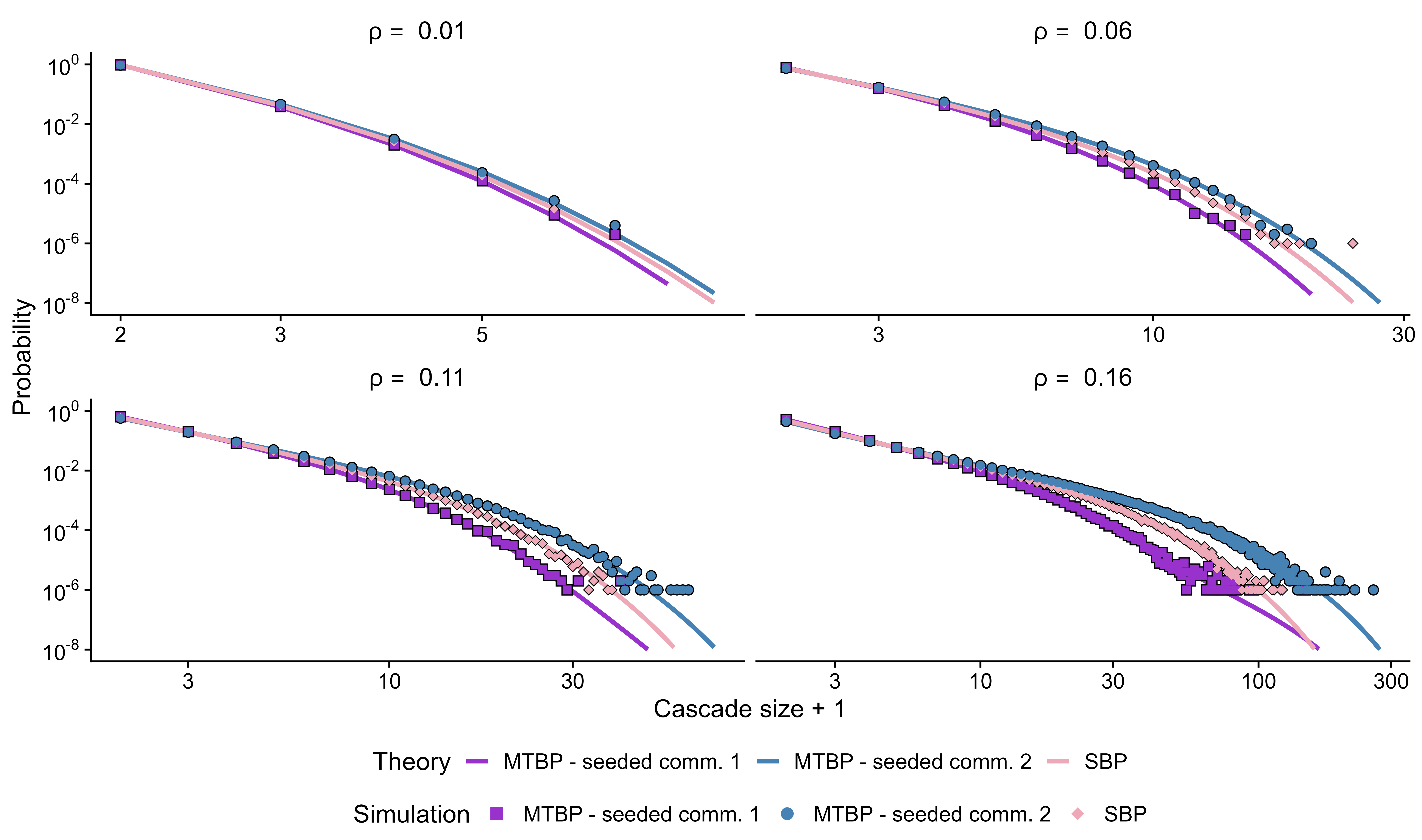}
    \caption{Cascade size across the entire network for a range of infection ($\rho$) parameters when seeded in different communities, leaving the cross-community connection probability, $p_e$, fixed at $0.02$. Each community has $5\,000$ nodes, totaling $10\,000$ nodes in the network. Community 1 has an internal average degree of $4$, while community 2 has an internal average degree of $5$. Note that the middle pink line is for the corresponding simple branching process (which does not take into account community structure).}
    \label{fig:appendix_pln_varying_rho}
\end{figure*}

We can note from Fig. \ref{fig:appendix_pln_varying_rho}, that again we have excellent agreement between our simulation and our theory curves. In Fig. \ref{fig:appendix_pln_varying_rho} it is hard this time to see any qualitative change in the shape of the cascade size distributions in community 1 or community 2 like we saw before.

However, we can still isolate any effect that we might see by comparing the cascade size distribution in community 1 when we allow inter community spreading ($p_e = 0.02$) and when we do not ($p_e = 0$) in Fig. \ref{fig:appendix_ln_effect_of_comm}a) and \ref{fig:appendix_ln_effect_of_comm}b). Again, it is hard to discern a difference with the simulations masking any patterns; However, if we focus on the theory curves with no simulations and examine a wider range of probabilities, shown in Fig. \ref{fig:appendix_ln_effect_of_comm}b), we can see a similar difference emerging. This difference occurs at a significantly smaller probability than previously, which means that even when community 1 does see a benefit from spreading to community 2, which has a slightly higher average degree, it only becomes pronounced at a very small probability, i.e., it would be rare to observe this effect for this network configuration. Again, we can isolate any effect that we might see by comparing the cascade size distribution in community 2 when we allow inter community spreading ($p_e = 0.02$) and when we do not ($p_e = 0$) in Fig. \ref{fig:appendix_ln_effect_of_comm}c). Like before, we cannot see a difference; which means that there is little qualitative difference when we allow spreading between communities in the case where we seed cascades in community 2.

\begin{figure*}[t]
    \centering
    \includegraphics[width=0.9\linewidth]{12.jpg}
    \caption{Cascade size across the entire network where $\rho$ is fixed to $0.16$. Purple and blue lines (MTBP) represent $p_e = 0.02$, where cross-community spread is possible, and pink lines (SBP) represent $p_e = 0$, where spread is confined to one community. a) Diffusion behavior when seeding in community 1 (which has an average degree of $\approx 4$). b) Diffusion behavior when seeding in community 1 (theory curves only to show the dynamics on cascades that spread further). c) Diffusion behavior when seeding in community 2 (which has an average degree of $\approx 5$). Here the blue and pink lines coincide. As before, each community has $5\,000$ nodes, totaling $10\,000$ nodes in the network. 
    }
    \label{fig:appendix_ln_effect_of_comm}
\end{figure*}

\clearpage
\newpage
\bibliographystyle{unsrt}
\bibliography{references}

\end{document}